\documentclass[11pt,showpacs,preprintnumbers,superscriptaddress,amsmath,amssymb,nofootinbib]{revtex4}
\usepackage{graphicx}
\usepackage{dcolumn}
\usepackage{bm}
\usepackage{amssymb}
\usepackage{amsmath}
\usepackage{epsfig}    
\usepackage{color}
\usepackage{slashed}
\usepackage{hhline}
\usepackage{mathrsfs}

\def\be{\begin{equation}}
\def\ee{\end{equation}}
\def\Mat3#1#2#3#4#5#6#7#8#9{
  \left(
  \begin{array}{ccc}
    #1 & #2 & #3 \\
    #4 & #5 & #6 \\
    #7 & #8 & #9 \\
  \end{array}
  \right) }
\newcommand{\bea}{\begin{eqnarray}}
\newcommand{\eea}{\end{eqnarray}}
\newcommand{\nn}{\nonumber}

\numberwithin{equation}{section}

\usepackage{slashed,color,graphicx}
\usepackage{hyperref}
\hypersetup{
  colorlinks = true,
  linkcolor = blue,
  citecolor = magenta
} 

\begin{document}

\title{ Neutrino Mass Model and Dark Matter with $Y=0$ Inert Triplet Scalar}
\preprint{KYUSHU-HET-273}

\author{ Shilpa Jangid}
\email{shilpa.jangid@apctp.org}
\affiliation{Asia Pacific Center for Theoretical Physics (APCTP) - Headquarters San 31,
Hyoja-dong, Nam-gu, Pohang 790-784, Korea}

\author{Keiko I. Nagao}
\email{nagao@ous.ac.jp}
\affiliation{Okayama University of Science, Faculty of Science,  Department of Physics, Ridaicho 1-1, Okayama, 700-0005, Japan}

\author{Hiroshi Okada}
\email{okada.hiroshi@phys.kyushu-u.ac.jp}
\affiliation{Department of Physics, Kyushu University, 744 Motooka, Nishi-ku, Fukuoka 819-0395, Japan}
\email{hiroshi3okada@htu.edu.cn}
\affiliation{Department of Physics, Henan Normal University, Xinxiang 453007, China}


\date{\today}

\begin{abstract}
We study a one-loop induced neutrino mass model with an inert isospin triplet scalar field of $Y=0$ and heavier isospin doublet vector-like leptons and singlet Majorana right-handed fermions. 
In addition to the neutrino mass matrix, We explain sizable scale of muon anomalous magnetic dipole moment $10^{-9}$ by introducing a singly-charged boson $S^\pm$.
We show numerical analysis of neutrino oscillation, lepton flavor violations, Z boson decays, and demonstrate our allowed regions in cases of normal and inverted hierarchies. We find the sizable scale of muon anomalous magnetic dipole moment for both cases.
Then, we move on to the discussion of dark matter candidates to satisfy the relic density where we have two candidates; fermionic dark matter and bosonic one. 
And, we classify four cases 
fermionic dark matter with normal and inverted hierarchies,
bosonic one with normal and inverted hierarchies
and search for each of the allowed points in the model. 
\end{abstract}
\maketitle
\newpage

\section{Introduction}
The mechanism of the neutrino mass matrix must be one of the biggest issues on how to be constructed or fixed,
because the neutrinos have specific natures, such as electrically neutral particles, and so tiny masses
which would not possess the other three fermion sectors in the standard model (SM).
Moreover, there exist some observables of CP Dirac phase, Majorana phases~\footnote{These phases appear only when the neutrinos are assumed to be Majorana particles. In this paper, however, we expect the neutrinos to be Majorana ones.}, and the neutrino masses, which would not still exactly be measured by current experiments yet. Thus, a neutrino model building depends on what kind of phenomenology one would like to explain or predict. For example, canonical seesaw~\cite{Yanagida:1979gs, Minkowski:1977sc, Mohapatra:1979ia} can explain the tiny neutrino masses by introducing heavier right-handed neutral fermions; ${\it e.g.}$ ${\cal O}(10^{15})$ GeV, that leads to successful explanation of the Baryon Asymmetry of the Universe via leptogenesis~\cite{Fukugita:1986hr}.
Then, alternative seesaw scenarios such as inverse seesaw~\cite{Mohapatra:1986bd, Wyler:1982dd} and linear seesaw~\cite{Wyler:1982dd, Akhmedov:1995ip, Akhmedov:1995vm} are developed to reduce the energy scale in which one more neutral fermions with different chirality are introduced.
Another example is that the neutrino masses are induced at loop level, and it is also a natural explanation for the smallness of the neutrinos.~\footnote{Even though there are a lot of models, we shall refer to a minimum model~\cite{Ma:2006km}.}
This model can often be realized at a lower energy scale, which can reach the current experimental energy $\sim$TeV. It implies that the model testability could be much better than the other scenarios. In addition, an intriguing dark matter (DM) candidate can be involved in since it interacts with the lepton sector. Therefore, it would be a flavorful (leptophillic) DM candidate, and it can be a clean signal in collider experiments.
In order to realize this model, we typically need to impose an additional symmetry (such as $Z_2$ symmetry) to stabilize the DM candidate and consider lepton flavor violations (LFVs), which give stringent constraints for the corresponding Yukawa couplings and masses.
Thus, we can discuss various phenomenologies within reach of current experimental energy and a vast alternative literature has been arisen along this line of idea. 

In this paper, we apply an isospin triplet boson with zero hypercharge $\Delta_0${, isospin doublet vector-like leptons $L'$ and right-handed Majorana neutral fermions $N_R$ to the neutrino mass matrix so that the neutrino masses are radiatively generated via $\Delta_0,\ L',\ N_R$}, imposing $Z_2$ symmetry to stabilize the DM candidate. 
As a result, we have two DM candidates; fermionic DM and bosonic one.
Especially, $\Delta_0$ has intriguing features as a dark matter (DM) candidate~\cite{Araki:2011hm, Jangid:2020qgo, Brdar:2013iea, Law:2013saa}.
At first, the neutral component of $\Delta_0$; $\delta_0$, is uniquely lighter than the singly-charged component; $\delta^\pm$, even though the mass difference is induced at not tree-level but radiative correction at one-loop level~\cite{Cirelli:2009uv} as follows:
\begin{align}
m_{\delta^\pm}=m_{\delta_0}+166\ {\rm MeV},
\end{align}
where $m_{\delta^\pm}$ and $m_{\delta_0}$ are respectively the mass of singly charged and neutral boson of $\Delta_0$. Thus, it is automatically a good DM candidate among the component of $\Delta_0$, if $\delta_0$ is stable.
%
The second feature is that the constraint of the S parameter is totally safe due to $Y=0$.
Moreover, there is no bound of direct detection of spin-independent scattering via the Z boson portal
even though there exists the bound via Higgs portal. It implies that this DM candidate still has allowed region for the constraints of direct detection bounds since contribution to the Z boson portal gives a much stronger bound than one to the Higgs boson portal.
{In addition, we explain the muon anomalous magnetic moment (muon $g-2$) by introducing a singly-charged boson $S^\pm$. But the boson also contributes to the Z boson decays at one-loop level that are precisely tested by experiments and we need to consider these constraints.}

This paper is organized as follows.
In Sec.~II, we show our model,  including the Higgs sector and neutral fermion sector.
Then, we have formulate the neutrino mass matrix at one-loop level,  LFVs, Z boson decays, and muon anomalous magnetic dipole moment.
And we have numerical analysis in which we firstly provide benchmark points (BPs) that have sizable muon anomalous magnetic dipole moment without conflict of any constraints such as LFVs and Z boson decays. Then, we show numerical analyses for cases of fermionic DM  and bosonic one respectively.
In Sec.~III, we analyze fermionic and bosonic DM candidates to explain the relic density for each case of NH and IH.
We conclude and discuss in Sec.~IV.


\section{ Model setup}
 \begin{widetext}
\begin{center} 
\begin{table}
\begin{tabular}{|c||c|c|c|c||c|c|c|}\hline\hline  
&\multicolumn{4}{c||}{Lepton Fields} & \multicolumn{3}{c|}{Scalar Fields} \\\hline
& ~$L_L$~ & ~$e_R^{}$~ & ~$N_R$~ & ~$L'$ ~ & ~$H$~  & ~$\Delta_0$~ & ~$S^{+}$ \\\hline 
$SU(2)_L$ & $\bm{2}$  & $\bm{1}$ & $\bm{1}$  & $\bm{2}$ & $\bm{2}$ & $\bm{3}$ & $\bm{1}$ \\\hline 
$U(1)_Y$ & $-\frac12$ & $-1$& $0$  & $-\frac{1}{2}$ & $\frac12$ & $0$ & $+1$  \\\hline
 $Z_2$ & $+$ & $+$   & $-$ & $-$ & $+$& $-$  & $-$  \\\hline
$\mathbb{L}$ & $1$   & $1$  & $1$ & $1$& $0$& $0$& $-2$ \\\hline\hline
\end{tabular}
\caption{Contents of fermion and scalar fields
and their charge assignments under $SU(2)_L\times U(1)_Y\times Z_2$, where $\mathbb{L}$ is the lepton number.}
\label{tab:1}
\end{table}
\end{center}
\end{widetext}

In this section, we review our model setup. 
As for the fermion sector, we introduce three heavy right-handed neutral fermions $N_R$ and three isospin doublet vector-like leptons $L'\equiv[n',e']$.
As for the boson sector, we add an isospin inert triplet $\Delta_0$ with zero hypercharge and an isospin singlet singly-charged boson $S^+$.
{Note here that $\Delta_0$ is a complex boson in general that plays a role in generating the neutrino mass matrix.
While $S^+$ contributes to the sizable muon $g-2$.}
The SM Higgs is denoted by $H$.
 All these new particles are imposed by $Z_2$ odd number and 
their contents and charges are summarized in Table~\ref{tab:1}.
Therefore, we have a fermionic DM candidate or bosonic one, as discussed later.
{Moreover, we impose lepton number $\mathbb{L}$ that is a global $U(1)$ symmetry and we allow the symmetry to be softly broken.}
The renormalizable Lagrangian in the lepton sector under these symmetries is found as
\begin{align}
-\mathcal{L}_{Y}
&=
y_{\ell_{ii}} \bar L_{L_i} H e_{R_i}  
+y_{\Delta_{ij}} \bar L_{L_i} \Delta_0 L'_{R_j}
{+y_{ij}  \bar L_{L_i}(i\tau_2) (L'^c_L)_j S^-}
\nn\\
&+ y_{S_{ij}} \bar N^c_{R_i} e_{R_j} S^+
+y_{N_{ij}} \bar L'_{L_i} (i\tau_2) H^* N_{R_j}
+ M_{R_{ii}} \bar N^c_{R_i} N_{R_i}  + M_{L_{ii}} \bar L'_{L_i} L'_{R_i} + {\rm h.c.}, 
\label{Eq:lag}
\end{align}
where $i=1-3$, $j=1-3$, $\tau_2$ is the second Pauli matrix. 
Without loss of generality, we work on the diagonal bases for $y_\ell, M_R, M_L$ {and $M_R$ is softly broken by two units of $\mathbb{L}$.}


\subsection{ Higgs potential}
The renormalizable Higgs potential under these symmetries is given by
\begin{align}
{\cal V}_{tri.}&=
\mu^2_{H} H^\dag H + {\mu^2_S} S^+ S^- + \mu^2_\Delta {\rm Tr}[\Delta_0^\dag\Delta_0]\nn\\
&+\lambda_H (H^\dag H)^2 + {\lambda_S}(S^+ S^-)^2 + \lambda_\Delta {\rm Tr}[(\Delta_0^\dag\Delta_0)^2]
+\lambda_{HS} (H^\dag H)(S^+ S^-)
\nn\\
&+\lambda_{H\Delta} (H^\dag H){\rm Tr}[\Delta_0^\dag\Delta_0]
+\lambda'_{H\Delta} \sum_{i=1}^3(H^\dag \tau_i H){\rm Tr}[\Delta_0^\dag\tau^i\Delta_0]
\label{Eq:pot1}, \\
{\cal V}_{non-tri.}&= m^2_\Delta {\rm Tr}[\Delta_0^2]
+{\rm h.c.},
\label{Eq:pot2}
\end{align}
where $\tau_i(i=1,2,3)$ being Pauli matrices, and ${\cal V}_{tri.}({\cal V}_{non-tri.})$ is trivial(nontrivial) potential. {Moreover, we forbid the term $H^T (i\tau_2) \Delta_0 H S^-$ because it dynamically violates the lepton number.~\footnote{{When this term is allowed, we have two new sources of muon $g-2$ and the neutrino mass matrix. The muon $g-2$ has no chiral suppression and we would expect to explain the muon $g-2$ via $y_\Delta$ and $y_S$. The new neutrino mass matrix is generated via $y_\Delta$ and $y$ at one-loop level.
This scenario is intriguing since $y_\Delta$ contributes to the neutrino mass matrix as well. However its value is of the order $10^{-13}$ in our numerical check that is much smaller than the current expected value $\sim10^{-9}$.}} Thus, we have no mixings among singly-charged bosons $S^\pm$ and $\delta_{1,2}^\pm$.}
${\cal V}_{non-tri.}$ is more important in our scenario
that leads us to the real DM particle since it gives the mass difference between the mass of CP even and odd particle, and the mass difference is directly connected to the nonzero neutrino mass matrix.  
Here, the scalar fields can be parameterized as 
\begin{align}
&H =\left[
\begin{array}{c}
w^+\\
\frac{v+h+iz}{\sqrt2}
\end{array}\right],\quad 
\Delta_0 =\left[
\begin{array}{cc}
\frac{\delta_R+i\delta_I}{\sqrt2} & \sqrt2 \delta^+_1\\
\sqrt2 \delta^-_2 & -\frac{\delta_R+i\delta_I}{\sqrt2}
\end{array}\right],
\label{component}
\end{align}
where $v~\simeq 246$ GeV is vacuum expectation value (VEV) of the SM Higgs, and $w^\pm$
and $z$ are respectively absorbed by the longitudinal component of $W^\pm$ and $Z$ boson in the SM.
{
The mass matrix for singly-charged bosons in terms of $\delta^\pm_1$ and $\delta^\pm_2$ is given by
\begin{align} 
\left[\begin{array}{cc}
 \mu^2_\Delta +\frac{v^2}2(\lambda_{H\Delta}-\lambda'_{H\Delta}) & m^2_\Delta \\
m^2_\Delta &  \mu^2_\Delta +\frac{v^2}2(\lambda_{H\Delta}+\lambda'_{H\Delta}) 
 \end{array}\right].
\end{align}
Here we assume $m_\Delta<<\mu_\Delta$ so that $\delta_2$ does not mix with the other singly-charged bosons. 
This assumption would be natural because $m_\Delta$ is expected to be small in order to explain the tiny neutrino masses. In fact, we only have solutions under $(m_\Delta/\mu_\Delta)^2\lesssim0.005$ in our numerical analysis to explain the neutrino oscillation data. Thus, we work on the mass eigenstates for these three singly-charged bosons.
%
%
 }
{The mass matrix for the CP-even bosons in basis of $(h,\delta_R)$ is given by
\begin{align} 
\left[\begin{array}{cc}
\lambda_H v^2 &0 \\
0 &  \mu_\Delta^2 +\frac{\lambda_{H\Delta} v^2}{2} +m^2_\Delta \end{array}\right].
\end{align}
The mass matrix for the CP-odd bosons in basis of $(z,\delta_I)$ is given by
\begin{align} 
\left[\begin{array}{cc}
0 &0 \\
0 &  \mu_\Delta^2 +\frac{\lambda_{H\Delta} v^2}{2} - m^2_\Delta \end{array}\right].
\end{align}
Note here that the massless $z$ is eaten by the Z boson in the SM and these neutral bosons do not mix each other. Therefore,}
the mass eigenvalues for $\delta_R,\delta_I$ are respectively given by
\begin{align}
m_R^2\equiv \mu_\Delta^2 +\frac{\lambda_{H\Delta} v^2}{2} +m^2_\Delta,\quad
m_I^2 \equiv \mu_\Delta^2 +\frac{\lambda_{H\Delta} v^2}{2}-m^2_\Delta.
\end{align}
%
 The mass square difference of $\delta_R$ and $\delta_I$; $m^2_\Delta$, plays an important role in a nonzero neutrino mass matrix as can be seen later~\cite{Ma:2006km}.

\subsection{ Neutral fermion mass matrix}
Since we have three by three block heavier neutral fermion mass matrix based on $\vec N = [N_R,n'_R,n'^C_L]$ originated from $N_R$, $L'_R$, and $L'_L$, we need to formulate their mass eigenvalues and eigenvectors.
After spontaneous symmetry breaking, the valid mass terms are given by
\begin{align}
m'_{ij} \bar n'_{L_i} N_{R_j} + M_{R_{ii}} \bar N^c_{R_i} N_{R_i}  + M_{L_{ii}} \bar L'_{L_i} L'_{R_i} + {\rm h.c.},
\end{align}
where $m'_{ij} (\equiv y_{N_{ij}} v/\sqrt2)$ has to be within perturbative limit which we set $m'_{ij}\lesssim \sqrt{2\pi}v$. 
Then, the mass matrix $M_N$ {in basis of $[N_R,n'_R, n'^C_L]^T$} is given by
\begin{align}
M_N=
\left[\begin{array}{ccc}M_R &0 & m'^T\\
0 & 0 & M^T_{L}\\
m' & M_{L} & 0 \\
  \end{array}
\right],
\end{align}
and diagonalized by $V^T_N M_N V_N = D_N$ and $\vec N = V_N \vec\psi$.
It suggests that $N_R = \sum_{a=1}^9(V_N)_{i,a} \psi_{R_a}$, $n'_R = \sum_{a=1}^9(V_N)_{i+3,a} \psi_{R_a}$,
and $n'^C_L = \sum_{a=1}^9(V_N)_{i+6,a} \psi_{R_a}$.

\subsection{ Neutrino mass matrix}
\begin{figure}[tb]
\begin{center}
\includegraphics[scale=0.22]{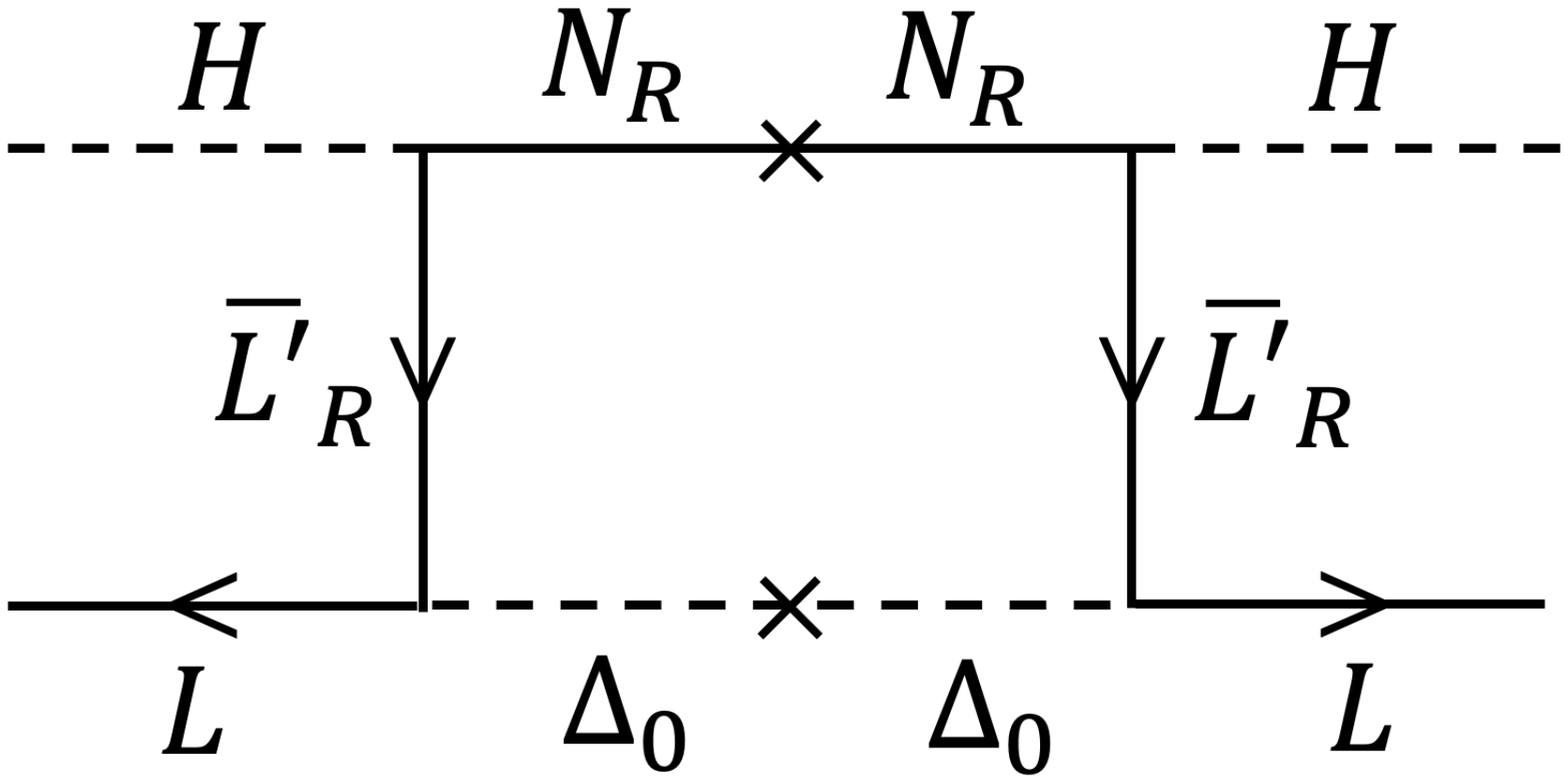}
\caption{Diagram for the neutrino mass.}
\label{fig:neutrinomass}
\end{center}
\end{figure}
Now that we have all the definitions for constructing the neutrino mass matrix, we discuss the neutrino sector.
{The neutrino mass matrix is induced at one-loop level in FIG.~\ref{fig:neutrinomass}. That is categorized by the T1-3 diagram in ref.~\cite{Bonnet:2012kz}.}
At first, we write down the relevant Lagrangian in terms of mass eigenstates of neutral heavier fermions as follows:
\begin{align}
\frac1{\sqrt2} (Y_\Delta)_{ia} \bar \nu_{L_i} \psi_{R_a} (\delta_R+i\delta_I)
+{\rm h.c.} ,
\label{Eq:lag-mass}
\end{align}
where we define $(Y_\Delta)_{ia}\equiv \sum_{j=1}^3 (y_\Delta)_{ij} (V_N)_{j+3,a}$ which is three by nine dimensionless matrix.
The resultant neutrino mass matrix is then given by
 \begin{align}
&(m_{\nu})_{ij}
=\sum_{a=1}^9 \frac{(Y_\Delta)_{ia} D_{N_a} (Y^T_\Delta)_{aj}}{2(4\pi)^2}
\left(
\frac{m^2_R}{m^2_R- D^2_{N_a}}\ln\left[\frac{m^2_R}{D^2_{N_a}}\right] 
- \frac{m^2_I}{m^2_I- D^2_{N_a}}\ln\left[\frac{m^2_I}{D^2_{N_a}}\right] 
 \right).
\end{align}
In order to apply the Casas-Ibarra parametrization~\cite{Casas:2001sr}, we redefine the above formula 
as follows:
 \begin{align}
(m_{\nu})_{ij} &\equiv\frac{1}{2(4\pi)^2}
(y_\Delta)_{i\alpha} 
\left[(V_N)_{\alpha+3 a}F_a (V_N^T)_{a\beta+3}\right] (y^T_\Delta)_{\beta j}
\equiv \frac{1}{2(4\pi)^2}(y_\Delta)_{i\alpha}  (\mu_M)_{\alpha\beta} (y^T_\Delta)_{\beta j} 
,\\
F_a&\equiv
\left(
\frac{m^2_R}{m^2_R- D^2_{N_a}}\ln\left[\frac{m^2_R}{D^2_{N_a}}\right] 
- \frac{m^2_I}{m^2_I- D^2_{N_a}}\ln\left[\frac{m^2_I}{D^2_{N_a}}\right] 
 \right).
\end{align}
Then, the symmetric parameter $\mu_M$ is uniquely decomposed by a lower unit triangular $R_N$ as $\mu_M=R_N R_N^T$~\cite{Nomura:2016run}.
Since $m_\nu$ is diagonalized by Pontecorvo-Maki-Nakagawa-Sakata mixing matrix $V_{\rm MNS}$~\cite{Maki:1962mu} as $D_\nu\equiv V_{\rm MNS}^T m_\nu V_{\rm MNS}$ due to the diagonal charged-lepton sector, we obtain the following Casas-Ibarra parametrization
\begin{align}
y_\Delta=\sqrt2(4\pi)V_{\rm MNS}^T D_\nu^{1/2} O_{mix} R_N^{-1}\lesssim
\sqrt{4\pi},\label{eq:neutcond}
\end{align}
where $O_{mix}$ is three by three orthogonal complex matrix; $O_{mix} O_{mix}^T=O_{mix}^T O_{mix} =1$ and perturbative limit be imposed for $y_\Delta$.
We impose $V_{\rm MNS}$ and $D_\nu$ be within the current experimental results (two Majorana phases) where we refer to Nufit 5.2~\cite{Esteban:2020cvm} without SK.
Although the typical cosmological upper bound is given by $\sum D_\nu\lesssim 120$ meV~\cite{Planck:2018vyg},
we adopt more relaxed cosmological upper bound $\sum D_\nu\lesssim 151$ meV~\cite{Vagnozzi:2017ovm}.
Here $\sum D_\nu\equiv D_{\nu_1}+D_{\nu_2}+D_{\nu_3}$
The effective mass for the neutrinoless double beta decay is defined by
\begin{align}
\langle m_{ee}\rangle=\left|\sum_{i=1}^3 D_{\nu_i} (V_{\rm MNS})_{ei}^2\right|,
\end{align}
where it is constrained by the current KamLAND-Zen data~\cite{KamLAND-Zen:2016pfg} and could be measured in its future experiment~\cite{KamLAND-Zen:2022tow}.
The upper bound is given by $\langle m_{ee}\rangle<(61-165)$ meV at 90 \% confidential level (CL).
%
{
Note here that we implicitly satisfy the LFV constraints via $y_\Delta$ in our numerical analysis. In fact, this constraints are not so strong unless the Yukawa couplings are order $10^{-3}$ and the running masses are the order 100 GeV.
}

\subsection{ Lepton Flavor Violations and $Z$ boson decays}
\label{lfv-lu}
In our model, there exist 
$\ell_b\to\ell_a \gamma$ processes and $Z$ boson decays; $Z\to\ell_a\bar\ell_b$, without chiral suppression and we have to satisfy these constraints in order to
explain the muon $g-2$.
These processes arise from the following terms
\begin{align}
(Y_S)_{aj} \bar\psi^C_{R_a} e_{R_j} S^+
{-
Y_{ia}\bar \ell_{L_i} \psi_{R_a} S^-}
+{\rm h.c.},
\label{Eq:lag-mass}
\end{align}
where $(Y_S)_{aj}\equiv \sum_{i=1}^3(V^T_N)_{ai} (y_S)_{ij}$ {and
$Y_{ia}\equiv \sum_{j=1}^3 y_{ij} (V_N)_{j+6,a}$}.
{\bf The branching ratio of ${\rm BR}(\ell_b\to\ell_a \gamma)$} is defined by
\begin{align}
{\rm BR}(\ell_i\to\ell_j \gamma)
=
\frac{48\pi^3 C_{ij}\alpha_{\rm em}}{{\rm G_F^2} m_i^2 }(|a_{R_{ij}}|^2+|a_{R_{ij}}|^2),
\end{align}
where $\alpha_{\rm em}\approx1/137$ is the fine-structure constant,
$C_{ij}\approx(1,0.1784,0.1736)$ for ($(ij)=(21),(31),(32)$), ${\rm G_F}\approx1.17\times 10^{-5}$ GeV$^{-2}$ is the Fermi constant.
$a_L$ and $a_R$ are respectively computed as
\begin{align}
a_{R_{ij}}&=
{
-Y_{ja} D_{N_a} (Y_S)_{ai}G(D_{N_a},m_{S}),
\quad
a_{L_{ij}}=
-(Y_S^\dag)_{ja} D_{N_a} Y^\dag_{ai} G(D_{N_a},m_{S}),
}
\\
%
G(m_1,m_2)& \equiv  \int dxdy dz\delta(x+y+z-1)\frac{y}{x m_1^2 + (y+z) m_2^2},
\end{align} 
{where $m_S$ is the mass eigenvalue of $S^\pm$.}
These branching ratios have to be compared with experimental values in Table~\ref{tab:Cif}.
Notice here that three body decay processes $\ell_i\to\ell_j\ell_k\bar \ell_\ell$ at the one-loop box type of diagrams are almost negligible comparing to the ${\rm BR}(\ell_i\to\ell_j \gamma)$ types of LFVs~\cite{Lindner:2016bgg}. Even though the upper bound of the branching ratio of $\mu\to e\gamma$ is recently updated by $3.1\times 10^{-13}$ in ref.~\cite{MEGII:2023ltw}, we will apply the previous upper bound in our numerical analysis.

\begin{table}[t]
\begin{tabular}{c|c|c|c} \hline
Process & $(ij)$ & Experimental bounds ($90\%$ CL) & References \\ \hline
$\mu^{+} \to e^{+} \gamma$ & $(21)$ &
	$\text{Br}(\mu \to e\gamma) < 4.2(3.1)\times 10^{-13}$  &\cite{MEG:2016leq}(\cite{MEGII:2023ltw}) \\
$\tau^{\pm} \to e^{\pm} \gamma$ & $(31)$ &
	$\text{Br}(\tau \to e\gamma) < 3.3 \times 10^{-8}$  & \cite{BaBar:2009hkt}\\
$\tau^{\pm} \to \mu^{\pm} \gamma$ & $(32)$ &
	$\text{Br}(\tau \to \mu\gamma) < 4.4 \times 10^{-8}$   & \cite{BaBar:2009hkt}\\ \hline
\end{tabular}
\caption{Summary of $\ell_i \to \ell_j \gamma$ process and the upper bound of experimental data, where upper bound of branching ratio of $\mu\to e\gamma$ is recently updated by $3.1\times 10^{-13}$ in ref.~\cite{MEGII:2023ltw}.}
\label{tab:Cif}
\end{table}

{

 \begin{figure}[tb]
\includegraphics[width=10cm]{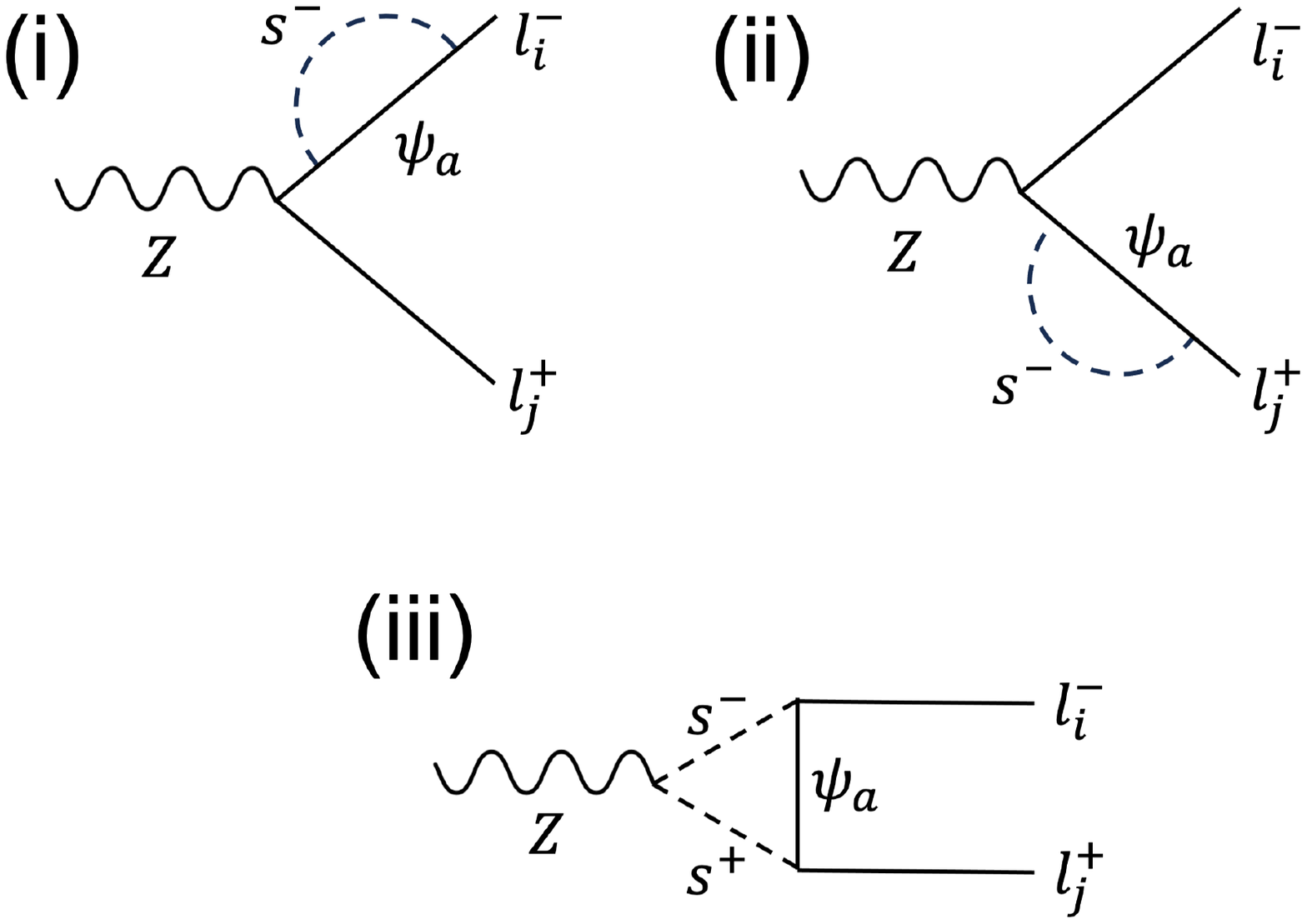} \ 
\caption{{
Diagrams for modifying interactions between $Z$ and SM charged leptons where dashed line indicate charged scalar bosons.}}
\label{fig:diagram3}
\end{figure}
{\bf The branching ratios of ${\rm BR}(Z\to\ell_a\bar\ell_b)$} receive modifications from the SM by our new contributions at one-loop level as shown in Figure~\ref{fig:diagram3}.
The modified widths of $Z$ boson are given by
\begin{equation}
\Gamma^{\rm SM+NP}_{Z \to \ell_i \overline{\ell_j} } = \frac{g_2^2}{24 \pi c_W^2} m_Z \left[ \left( - \frac12 + s^2_W \right)^2 \left| \delta_{ij} + \epsilon_{L_{ij}} \right|^2 
+ s_W^4 \left|\delta_{ij} + \epsilon_{R_{ij}} \right|^2 \right].
\end{equation}
$\delta_{ij}$ is the SM contributions
where $\delta_{ij}=1$ for $i=j$ otherwise $\delta_{ij}=0$.
$\epsilon_{L,R}$ is the new ones and given as follows:
\begin{align}
&\epsilon_{L_{ij}}\simeq
\frac{D_{N_a}}{(4\pi)^2}\left(m_j Y_{ia} (Y_S)_{aj}+m_i (Y^{\dag}_S)_{ia} Y^{\dag}_{aj} \right)H_a \nn\\
&-\frac{s_W^2}{\left(-\frac12+s_W^2\right)}\frac{D_{N_a}}{(4\pi)^2}
\left(m_j Y_{ia} (Y_S)_{aj} \Pi_{1_a} +m_i (Y^{\dag}_S)_{ia} Y^{\dag}_{aj} \Pi_{2_a} \right)\nn\\
&-\frac{1}{(4\pi)^2}
\left[
Y_{ia}Y^\dag_{aj} 
\left(-\frac{m_i^2\Pi_{3_{ai}}-m_j^2\Pi_{3_{aj}}}{m^2_i - m^2_j}
+\Pi_{4_a}
\right) 
+ m_i m_j Y^\dag_{S_{ia}} Y_{S_{aj}} H_a
\right]
,\label{eq:eL}\\
&\epsilon_{R_{ij}}\simeq
\frac{D_{N_a}}{(4\pi)^2}\left(m_i Y_{ia} (Y_S)_{aj}+m_j (Y^{\dag}_S)_{ia} Y^{\dag}_{aj} \right)H_a \nn\\
&-\frac{s_W^2}{\left(-\frac12+s_W^2\right)}\frac{D_{N_a}}{(4\pi)^2}
\left(m_i Y_{ia} (Y_S)_{aj} \Pi_{2_a}+m_j (Y^{\dag}_S)_{ia} Y^{\dag}_{aj} \Pi_{1_a}\right)\nn\\
&-\frac{1}{(4\pi)^2}
\left[Y^\dag_{S_{ia}} Y_{S_{aj}} 
\left(-\frac{m_i^2\Pi_{3_{ai}}-m_j^2\Pi_{3_{aj}}}{m^2_i - m^2_j}
+\Pi_{4_a}
\right)
+m_i m_j Y_{ia}Y^\dag_{aj} H_a
\right]\label{eq:eR},
\end{align}
%
where the loop functions $H, \Pi_{1,2}$ are computed by
\begin{align}
H_a&\simeq \frac{2 D_{N_a}^6+3D_{N_a}^4m_S^2-6 D_{N_a}^2 m_S^4 + m_S^6+6 D_{N_a}^4m_S^2\ln\left(\frac{m_S^2}{D_{N_a}^2}\right)}{6(D_{N_a}^2-m_S^2)^4},\\
\Pi_{1_a}&\simeq 
\int[dx]_3 \frac{1-2z}{x D_{N_a}^2 +(y+z) m_S^2-z(1+x-z) m^2_Z},\\
\Pi_{2_a}&\simeq 
\int[dx]_3 \frac{1+2x-2z}{x D_{N_a}^2 +(y+z) m_S^2-z(1+x-z) m^2_Z},\\
\Pi_{3_{ai}}&\simeq 
\int[dx]_2 y \ln[x D_{N_a}^2 +y m_S^2 + (y^2-y)m^2_i],\\
\Pi_{4_a}&\simeq 
\int[dx]_3 \ln[x D_{N_a}^2 +(y+z) m_S^2-z(1+x-z) m^2_Z],
\end{align}
$[dx]_y\equiv dxdy\delta(1-x-y)$ and $[dx]_3\equiv dxdydz\delta(1-x-y-z)$.

Note that we have flavor non-conserving $Z$ boson decays from new physics contributions. Then we impose the constraints~\cite{ParticleDataGroup:2020ssz}:
\begin{align}
& {\rm BR}(Z \to e^\pm \mu^\mp) \leq 7.5 \times 10^{-7}, \quad {\rm BR}(Z \to e^\pm \tau^\mp) \leq 9.8 \times 10^{-6}, \nonumber \\
& {\rm BR}(Z \to \mu^\pm \tau^\mp) \leq 1.2 \times 10^{-5}.
\end{align}
For flavor conserving modes, we define deviation from the SM prediction by
\begin{equation}
\Delta {\rm BR} (Z \to \ell_i \overline{\ell_i})  = \frac{\Gamma^{\rm SM+NP}(Z \to \ell_i \overline{\ell_i} ) - \Gamma^{\rm SM}(Z \to \ell_i \overline{\ell_i} ) }{\Gamma_Z^{\rm tot}},
\end{equation}
where $\Gamma_Z^{\rm tot} = 2.4952 \pm 0.0023$ GeV is the total decay width of $Z$ boson. 
We impose the current bounds on the deviation for the lepton flavor conserving $Z$ boson decays which are given by~\cite{ParticleDataGroup:2020ssz}
\begin{align}
& |\Delta {\rm BR}(Z \to e^+ e^-)| \leq 4.2 \times 10^{-5}, \quad |\Delta {\rm BR}(Z \to \mu^+ \mu^-)| \leq 6.6 \times 10^{-5}, \nonumber \\
& |\Delta {\rm BR}(Z \to \tau^+ \tau^-)| \leq 1.2 \times 10^{-5} .
\end{align}
Notice that we do not consider $Z \to \nu \bar \nu$ decay mode since new physics effect is smaller and the experimental constraints are also less significant.

}

\subsection{Muon anomalous magnetic moment}
In using the same formula as LFVs, we also describe a formula for muon $g-2$; $\Delta a_\mu$, as follows:
\begin{align}
\Delta a_\mu\approx -\frac{m_\mu}{2(4\pi)^2}(a_{R_{22}}+a_{L_{22}}).
\end{align}
The above formula should be within the current experimental value whose new results on the muon $g-2$ were recently published by the E989 collaboration at Fermilab \cite{Muong-2:2021ojo}: 
\begin{align}
a^{\rm FNAL}_\mu =116592040(54) \times 10^{-11}.
\label{exp_dmu}
\end{align}
Combined with the previous BNL result, the result indicates that the muon $g-2$ deviates from the SM prediction by 4.2$\sigma$ level
and its deviation from the SM prediction at 1$\sigma$ is given by
\begin{align}
\Delta a^{\rm exp}_\mu =
a^{\rm FNAL}_\mu -  a_\mu^{\rm SM}= 
 (25.1\pm 5.9)\times 10^{-10} ,
\label{exp_dmu}
\end{align}
where $\Delta a^{\rm exp}_\mu$ is space of explanation by the new physics.


\subsection{Numerical analysis}
\label{subsec:numericalanalysis}
Here we show our allowed space to satisfy the neutrino oscillation data in the case of NH and IH, and LFVs and discuss whether our result would reach the sizable value of muon $g-2$.
We randomly scan our input absolute values in range $[0.01,\sqrt{4\pi}]$ for nine complex values of $y_S$ and $y_{21,22,23}$~\footnote{We set these three components to be nonzero in order to obtain sizable muon $g-2$. As a result, the branching ratio of $\tau\to e\gamma$ is zero.},
$|m'_{}|\lesssim \sqrt{2\pi}v$, $0\le |M_R|\lesssim 10^5$ GeV, and $100{\rm GeV}\le |M_{L'}|\lesssim 10^5$GeV. 
We assume that $m_I^2=m_R^2+2 m_\Delta^2$, and 
with $0\lesssim m_\Delta^2 \le (0.1 v)^2$ and $10^2{\rm GeV}\le (m_R,m_{S})\le10^{4}$GeV
that would be an appropriate hypothesis to satisfy the oblique parameters and LEP bounds.
{
If we do not consider the DM candidate, we obtain the maximum value of muon $g-2$ $8.22\times10^{-10}$ in case of NH and $1.12\times10^{-9}$ in case of IH
satisfying all the constraints of LFVs and Z boson decays. We summarize the bench mark points to satisfy this value for NH in Table~\ref{tab:nhbp} and IH in Table~\ref{tab:inbp}.
}

Furthermore, we classify FDM candidate or BDM one imposing the following conditions.
In case of FDM, we have to impose $D_{N_1}< m_R,m_I,m_{S}$ and $0.9\le |(V_N)_{11}|$
where the second condition implies DM is the lightest $N_R$ dominant.
Notice here that the DM is ruled out when DM is $n'_{L/R}$ dominant because of the Z boson portal in direct detection searches.
In case of BDM, assuming real inert boson $\delta_R$, we have to impose $m_R<D_{N_1},m_I,m_{S}$.
Note that the relic density of DM will be taken into account in the next section.

\begin{table}[h]
	\centering
	\begin{tabular}{|c|c|c|} \hline 
			\rule[14pt]{0pt}{0pt}
 		&  NH  \\  \hline
			\rule[14pt]{0pt}{0pt}
		$\Delta a_\mu$ & $8.22\times10^{-10}$       \\ \hline
		\rule[14pt]{0pt}{0pt}
%
		$[|\Delta {\rm BR}(Z \to e^+ e^-)|, |\Delta {\rm BR}(Z \to \mu^+ \mu^-)|,|\Delta {\rm BR}(Z \to \tau^+ \tau^-)|]$ & $[2.4\times10^{-10},5.8\times10^{-7},3.6\times10^{-8}]$   \\ \hline
		\rule[14pt]{0pt}{0pt}
		$[{\rm BR}(Z \to e^\pm \mu^\mp),{\rm BR}(Z \to e^\pm \tau^\mp), {\rm BR}(Z \to \mu^\pm \tau^\mp)] $ & $[7.8\times10^{-16},4.1\times10^{-17},1.8\times10^{-15}]$     \\ \hline
		\rule[14pt]{0pt}{0pt}
				$[\mu^{+} \to e^{+} \gamma, \tau^{+} \to e^{+} \gamma,\tau^{+} \to \mu^{+} \gamma]$ & $[3.2\times10^{-13}, 0.0, 4.38\times10^{-8}]$     \\ \hline
		\rule[14pt]{0pt}{0pt}
				$M_R/{\rm GeV}$ & $\left[\begin{array}{ccc}118 + 1077 i &0 & 0\\
0 & -19.4 + 1.43 i & 0 \\
0 & 0 & -12.1 + 306 i \\
  \end{array}
\right]$     \\ \hline
		\rule[14pt]{0pt}{0pt}
		$m'/{\rm GeV}$ & 
  $\left[\begin{array}{ccc}158 - 345 i & 5.01 + 1.14 i& 70.0 + 129 i\\
48.0 + 44.3 i& 5.00 - 96.0 i& -97.5 - 28.8 i \\
41.0 - 47.4 i& 316 + 489 i& 2.45 + 1.56 i \\
  \end{array}
\right]$     \\ \hline
		\rule[14pt]{0pt}{0pt}
		$M_{L'}/{\rm GeV}$ & $\left[\begin{array}{ccc}11180 - 34692 i &0 & 0\\
0 & 931 + 13248 i & 0 \\
0 & 0 & 9345 + 136.5 i \\
  \end{array}
\right]$    \\ \hline
		\rule[14pt]{0pt}{0pt}
		$y_S$ & $\left[\begin{array}{ccc}
-0.018 + 0.0079 i& -0.0026 - 0.018 i& 0.25 + 
 0.19 i \\
-0.0048 + 0.023 i& 1.2 - 1.8 i& -0.0095 + 
 0.014 i \\
-0.028 - 0.021 i& 0.046 + 0.12 i& 0.31 - 0.37 i \\
  \end{array}
\right]$    \\ \hline
		$y$ & $\left[\begin{array}{ccc}
0& 0& 0 \\
-0.00083 + 0.00041 i& 0.012 - 0.50 i& -0.025 + 
 0.046 i \\
0& 0& 0 \\
  \end{array}
\right]$    \\ \hline
		\rule[14pt]{0pt}{0pt}
		$[m_S, m_{\delta_R}, m_{\delta_I},m_{\delta_{1,2}}]/{\rm GeV}$  &  $[1672.36,1484.7, 1484.77,1484.7] $   \\ \hline
		\rule[14pt]{0pt}{0pt}
		$[\alpha,\beta,\gamma]$  &  $[-0.49-0.018i,-1.39-0.33i, -2.9+0.089 i] $   \\ \hline
		\rule[14pt]{0pt}{0pt}
		$[\alpha_{21},\,\alpha_{31}]$ &  $[183^\circ,\, 285^\circ]$   \\ \hline
		\rule[14pt]{0pt}{0pt}
		$[D_{\nu_1}, \sum D_{\nu_i}]/{\rm meV}$ &  $[0.012, 58.7]$      \\ \hline
	\end{tabular}
	\caption{Numerical benchmark point in case of NH to get the maximum muon $g-2$ without conflict of any constraints. Here, $\alpha,\beta,\gamma$ represent three mixing angles of $O_{mix}$, and $\alpha_{21,31}$ are Majorana phases.}
	\label{tab:nhbp}
\end{table}

\begin{table}[h]
	\centering
	\begin{tabular}{|c|c|c|} \hline 
			\rule[14pt]{0pt}{0pt}
 		&  IH  \\  \hline
			\rule[14pt]{0pt}{0pt}
		$\Delta a_\mu$ & $1.1\times10^{-9}$       \\ \hline
		\rule[14pt]{0pt}{0pt}
%
		$[|\Delta {\rm BR}(Z \to e^+ e^-)|, |\Delta {\rm BR}(Z \to \mu^+ \mu^-)|,|\Delta {\rm BR}(Z \to \tau^+ \tau^-)|]$ & $[3.5\times10^{-11},1.3\times10^{-7},5.3\times10^{-10}]$   \\ \hline
		\rule[14pt]{0pt}{0pt}
		$[{\rm BR}(Z \to e^\pm \mu^\mp),{\rm BR}(Z \to e^\pm \tau^\mp), {\rm BR}(Z \to \mu^\pm \tau^\mp)] $ & $[5.1\times10^{-19},5.7\times10^{-21},4.4\times10^{-16}]$     \\ \hline
		\rule[14pt]{0pt}{0pt}
				$[\mu^{+} \to e^{+} \gamma, \tau^{+} \to e^{+} \gamma,\tau^{+} \to \mu^{+} \gamma]$ & $[2.0\times10^{-13}, 0.0, 1.8\times10^{-8}]$     \\ \hline
		\rule[14pt]{0pt}{0pt}
				$M_R/{\rm GeV}$ & $\left[\begin{array}{ccc}-3986 + 265 i &0 & 0\\
0 & -4062 - 3878 i & 0 \\
0 & 0 & 14.5 + 9.89 i \\
  \end{array}
\right]$     \\ \hline
		\rule[14pt]{0pt}{0pt}
		$m'/{\rm GeV}$ & 
  $\left[\begin{array}{ccc}-19.6 - 34.6 i & 1.31 + 5.78 i& 33.9 + 1.04 i\\
-112 + 23.5 i & -127 - 139 i & -14.1 + 0.538 i \\
-281 - 105 I& 35.8 + 33.9 i& -3.20 + 2.84 i  \\
  \end{array}
\right]$     \\ \hline
		\rule[14pt]{0pt}{0pt}
		$M_{L'}/{\rm GeV}$ & $\left[\begin{array}{ccc}-1876 - 5168 i &0 & 0\\
0 & -3334 - 4463 i & 0 \\
0 & 0 & 77.4 - 2007 i \\
  \end{array}
\right]$    \\ \hline
		\rule[14pt]{0pt}{0pt}
		$y_S$ & $\left[\begin{array}{ccc}
0.017 - 0.0042 i & -3.0 + 0.99 i& -0.11 - 0.087 i \\
0.041 - 0.026 i& 0.90 - 0.69 i& 0.042 - 0.10 i \\
-0.020 - 0.011 i& 0.014 + 0.085 i& 0.056 - 0.079 i \\
  \end{array}
\right]$    \\ \hline
		$y$ & $\left[\begin{array}{ccc}
0& 0& 0 \\
-0.027 + 0.43 i& -0.0022 - 0.11 i& 0.17 + 0.017 i \\
0& 0& 0 \\
  \end{array}
\right]$    \\ \hline
		\rule[14pt]{0pt}{0pt}
		$[m_S, m_{\delta_R}, m_{\delta_I},m_{\delta_{1,2}}]/{\rm GeV}$  &  $[4139,126, 127, 128] $   \\ \hline
		\rule[14pt]{0pt}{0pt}
		$[\alpha,\beta,\gamma]$  &  $[-2.4 + 0.092 i,1.0 + 0.085 i, 1.1 - 1.7 i] $   \\ \hline
		\rule[14pt]{0pt}{0pt}
		$[\alpha_{21},\,\alpha_{31}]$ &  $[241^\circ,\, 245^\circ]$   \\ \hline
		\rule[14pt]{0pt}{0pt}
		$[D_{\nu_3}, \sum D_{\nu_i}]/{\rm meV}$ &  $[57.9, 147]$      \\ \hline
	\end{tabular}
	\caption{Numerical benchmark point in case of NH to get the maximum muon $g-2$ without conflict of any constraints. Here, $\alpha,\beta,\gamma$ represent three mixing angles of $O_{mix}$, and $\alpha_{21,31}$ are Majorana phases.}
	\label{tab:inbp}
\end{table}

\subsubsection{NH in case of FDM}

\begin{figure}[tb]
\begin{center}
\includegraphics[scale=0.22]{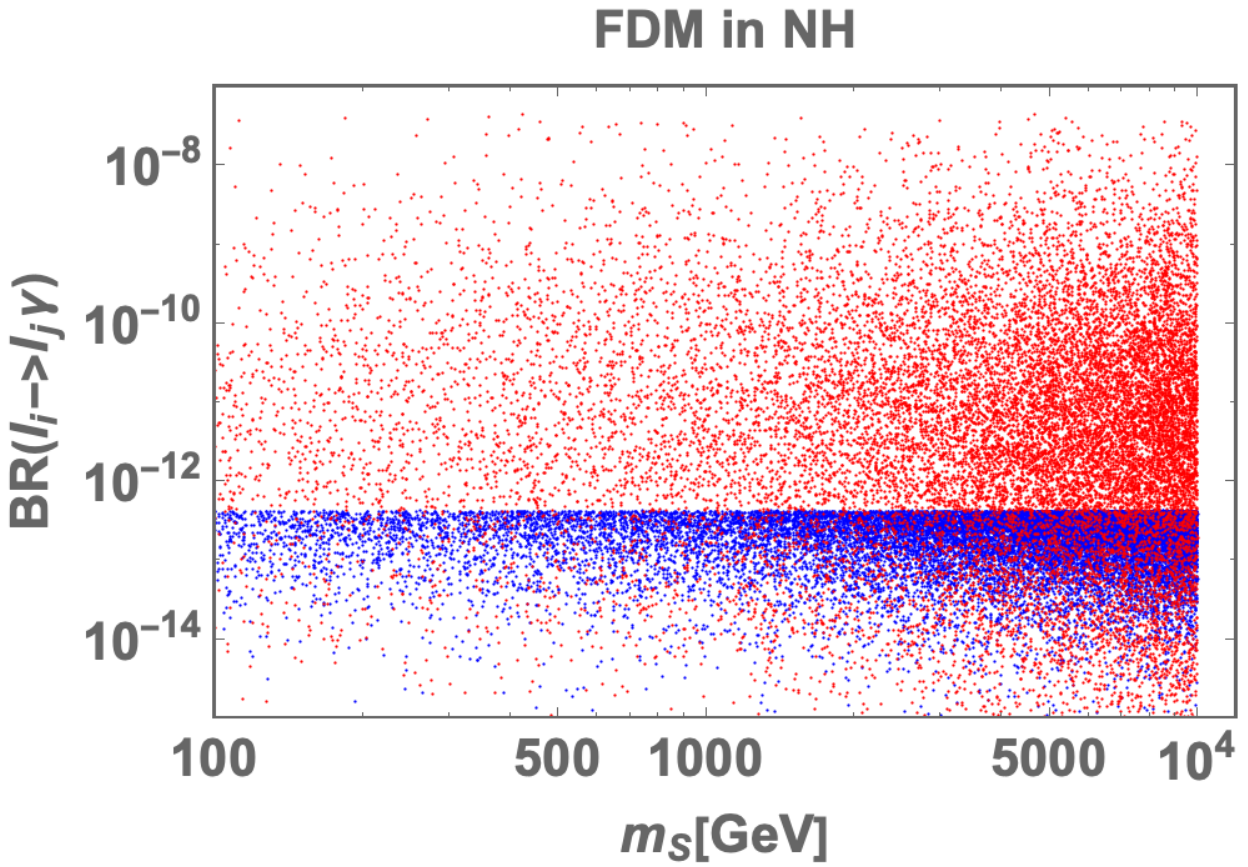}
\includegraphics[scale=0.22]{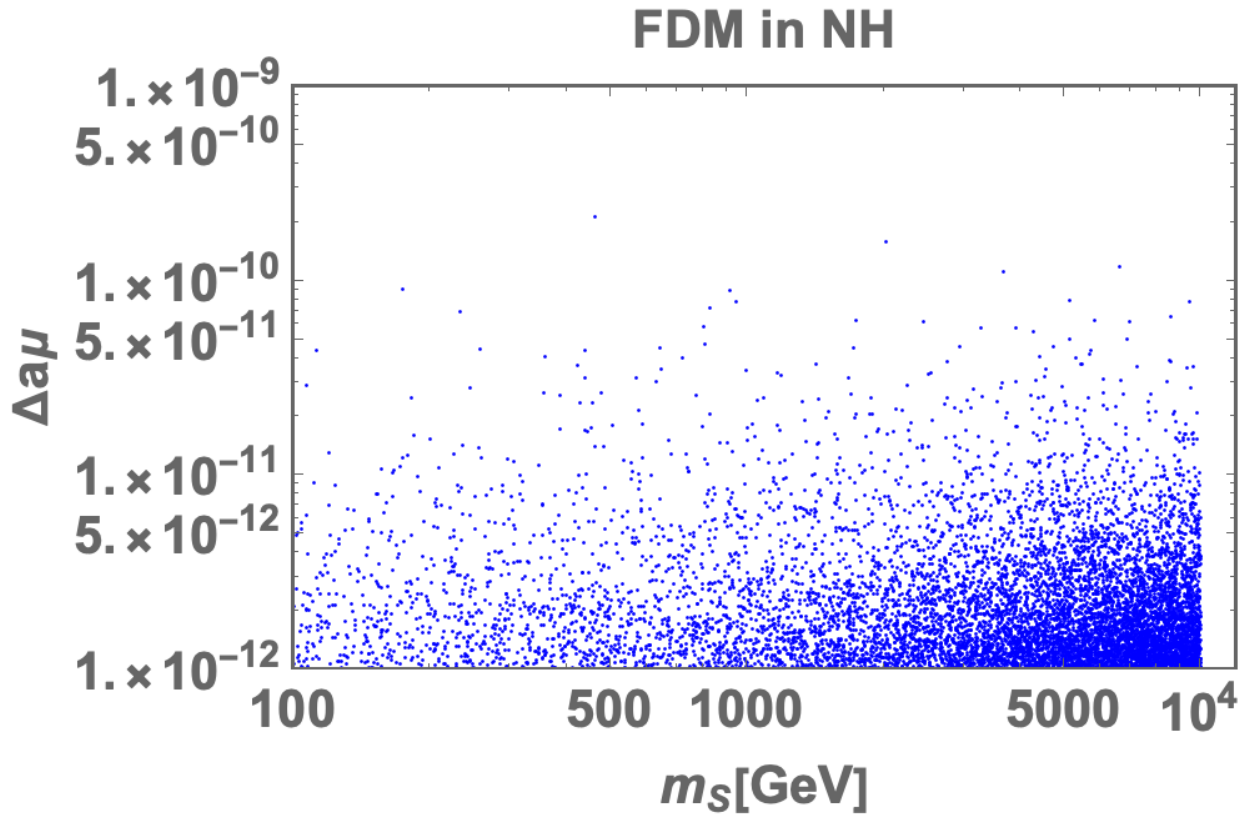}
\includegraphics[scale=0.22]{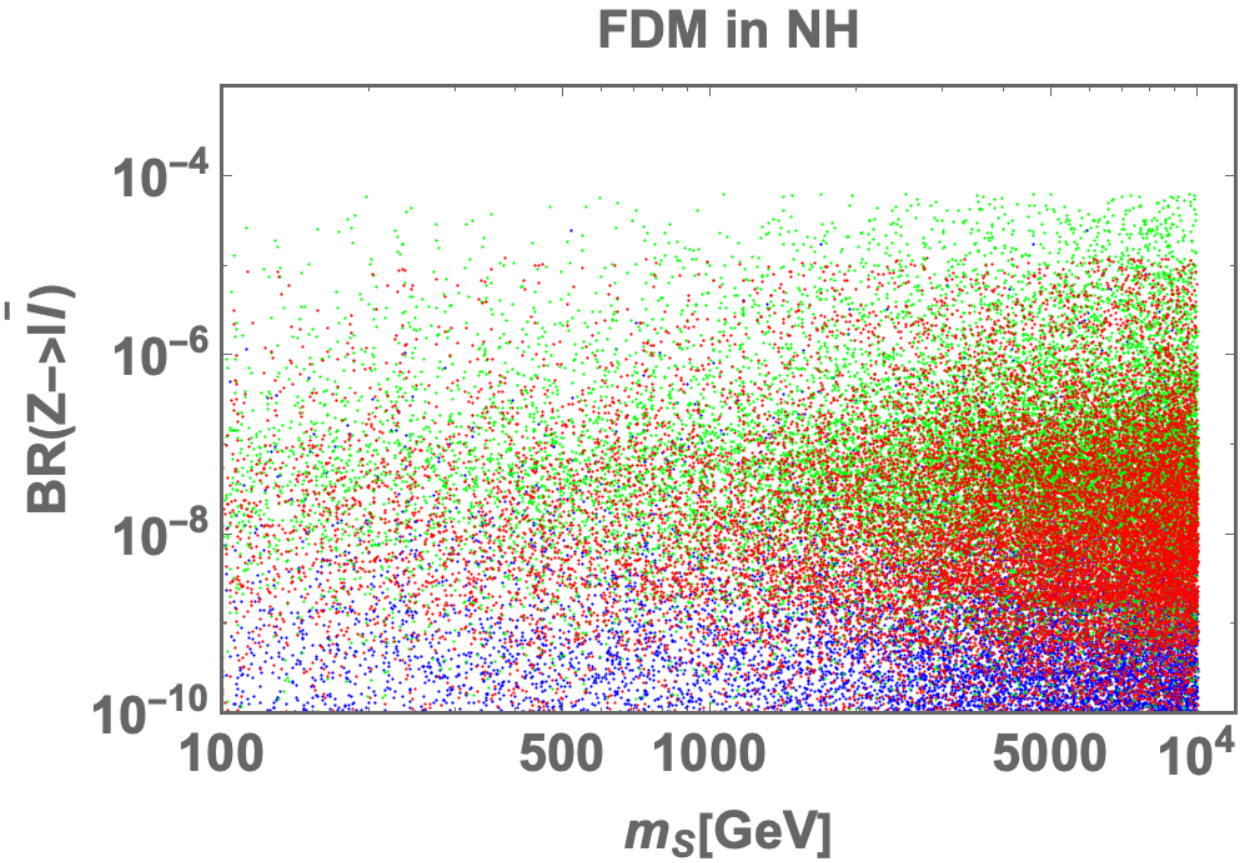}
\caption{Numerical results for the case of FDM candidate with NH: The left side figure represents the scattering plot of LFVs in terms of $m_S$ where the red points are for $\tau\to\mu\gamma$ and the blue ones $\mu\to e\gamma$.
The middle one represents the scattering plot of muon $g-2$ in terms of $m_S$. The order of muon $g-2$ reaches $10^{-10}$ that is smaller than the expected value by one order magnitude. 
The right one shows the scattering plot of BR($Z\to\ell\bar\ell$) in terms of $m_S$ where ${\rm BR}(Z\to e\bar e)$ is blue, ${\rm BR}(Z\to \mu\bar\mu)$ green and ${\rm BR}(Z\to \tau\bar\tau)$ red.}
\label{fig:fdmnh}
\end{center}
\end{figure}
%
In Figure~\ref{fig:fdmnh}, we show numerical results for the case of FDM candidate with NH
where we have obtained 18,752 allowed plots to satisfy the LFVs and Z boson decays.
The left side figure represents the scattering plot of LFVs in terms of $m_S$ where the red points are for $\tau\to\mu\gamma$ and the blue ones $\mu\to e\gamma$.
The middle one represents the scattering plot of muon $g-2$ in terms of $m_S$. The order of muon $g-2$ reaches $10^{-10}$.
The right one shows the scattering plot of BR($Z\to\ell\bar\ell$) in terms of $m_S$ where ${\rm BR}(Z\to e\bar e)$ is blue, ${\rm BR}(Z\to \mu\bar\mu)$ green and ${\rm BR}(Z\to \tau\bar\tau)$ red.
The right one implies that many plots are located in places near each the experimental limit of this Z boson decays that would be well-tested by future experiments.
On the other hand, we obtained the maximum values of flavor non-conserving Z boson decays in our numerical analysis as follows:
\begin{align}
& {\rm BR}(Z \to e^\pm \mu^\mp)^{max} \approx 4.3 \times 10^{-8}, \quad {\rm BR}(Z \to e^\pm \tau^\mp)^{max} \approx 1.6 \times 10^{-7}, \nonumber \\
& {\rm BR}(Z \to \mu^\pm \tau^\mp)^{max} \approx 1.0 \times 10^{-6},
\end{align}
which are smaller than the experimental constraints by 1 or 2 order of magnitudes.

\subsubsection{NH in case of BDM}
\begin{figure}[tb]
\begin{center}
\includegraphics[scale=0.22]{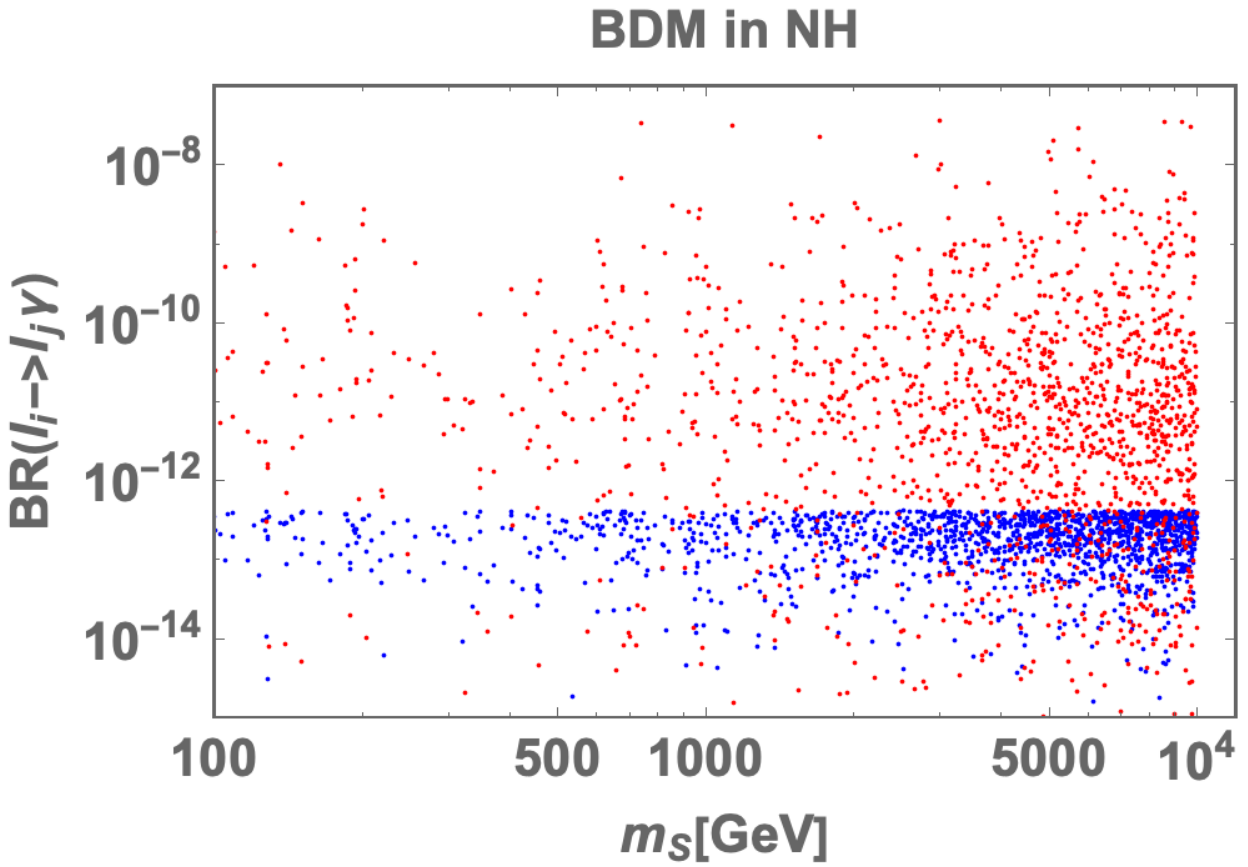}
\includegraphics[scale=0.22]{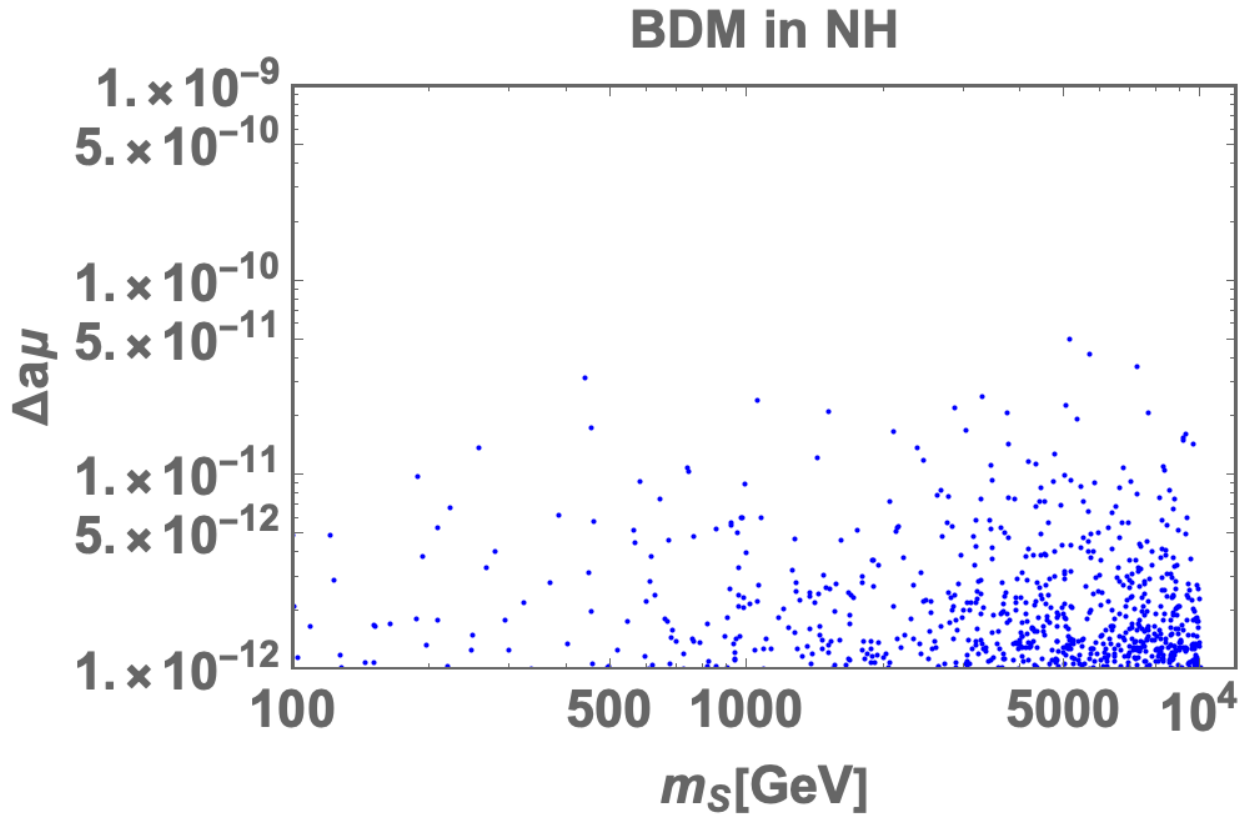}
\includegraphics[scale=0.22]{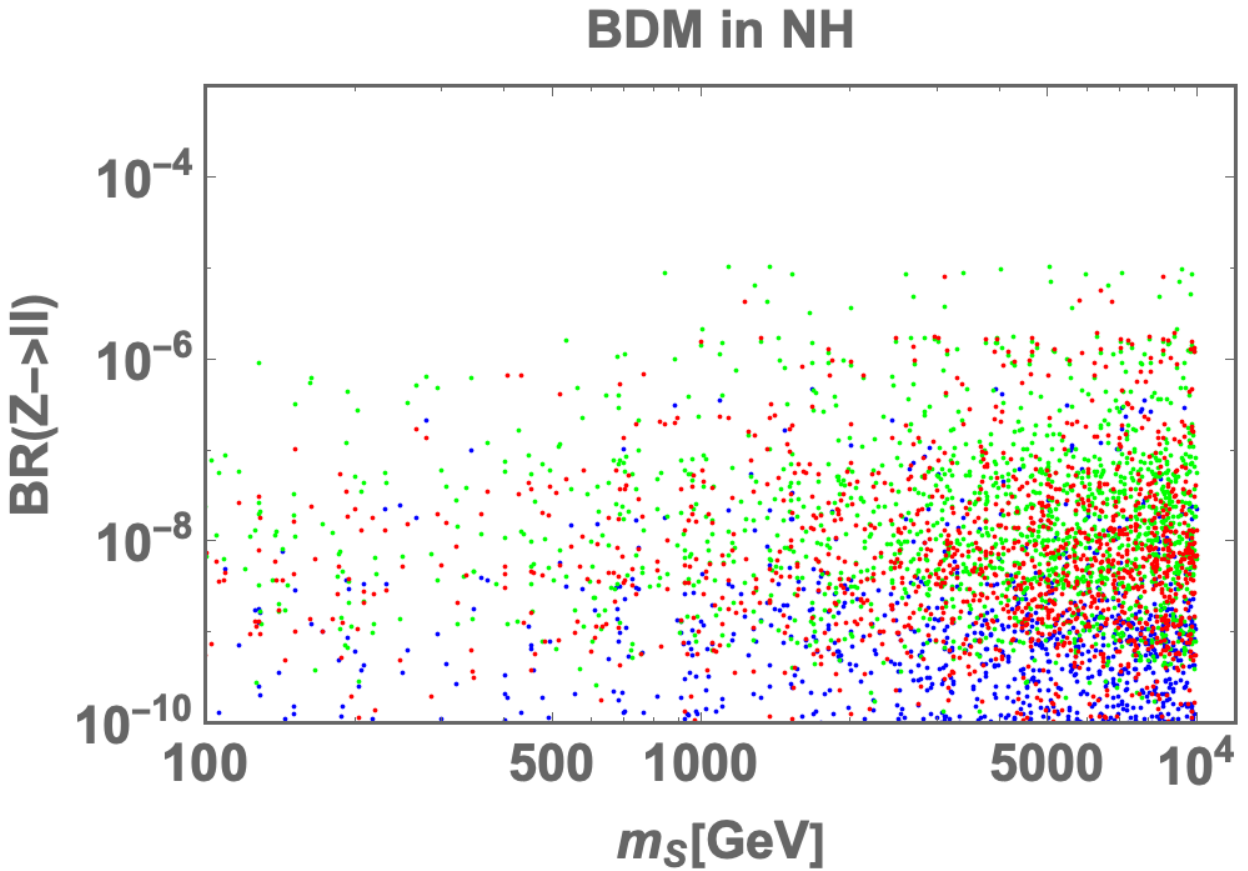}
\caption{Numerical results for the case of BDM candidate with NH where all the figure-legends are the same as the NH one. }
\label{fig:bdmnh}
\end{center}
\end{figure}
%
In Figure~\ref{fig:bdmnh}, we show numerical results for the case of BDM candidate with NH
where we have obtained 1,580 allowed plots to satisfy the LFVs and Z boson decays.
The left side figure represents the scattering plot of LFVs in terms of $m_S$ where the red points are for $\tau\to\mu\gamma$ and the blue ones $\mu\to e\gamma$.
The middle one represents the scattering plot of muon $g-2$ in terms of $m_S$. The order of muon $g-2$ reaches $5\times10^{-11}$ that is a little smaller than the case of FDM.
The right one shows the scattering plot of BR($Z\to\ell\bar\ell$) in terms of $m_S$ where ${\rm BR}(Z\to e\bar e)$ is blue, ${\rm BR}(Z\to \mu\bar\mu)$ green and ${\rm BR}((Z\to \tau\bar\tau))$ red.
The right one implies that many plots are located in places near each the experimental limit of this Z boson decays that would be well-tested by future experiments.
On the other hand, we obtained the maximum values of flavor non-conserving Z boson decays in our numerical analysis as follows:
\begin{align}
& {\rm BR}(Z \to e^\pm \mu^\mp)^{max} \approx 9.4 \times 10^{-12}, \quad {\rm BR}(Z \to e^\pm \tau^\mp)^{max} \approx 3.0 \times 10^{-12}, \nonumber \\
& {\rm BR}(Z \to \mu^\pm \tau^\mp)^{max} \approx 7.0 \times 10^{-11},
\end{align}
which are much smaller than the experimental constraints.

\subsubsection{IH in case of FDM}
\begin{figure}[tb]
\begin{center}
\includegraphics[scale=0.22]{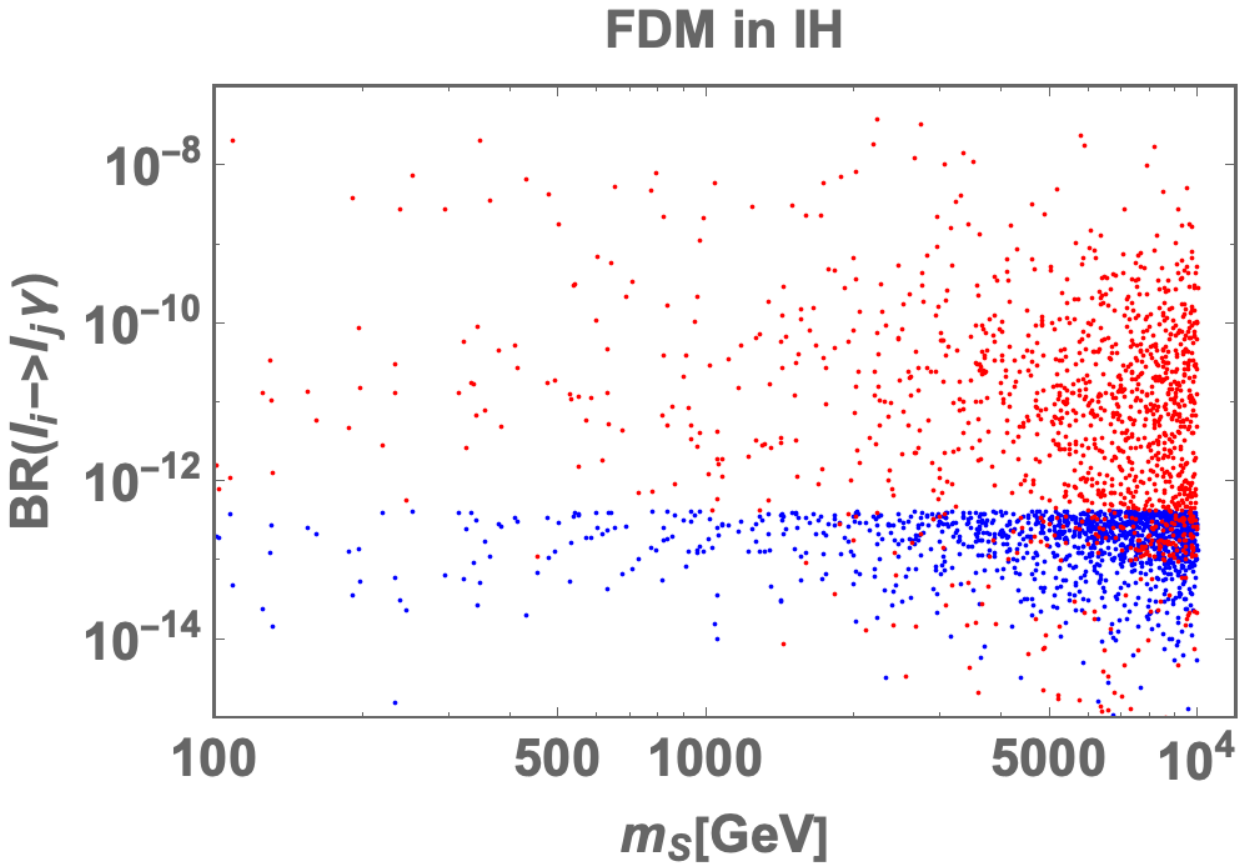}
\includegraphics[scale=0.22]{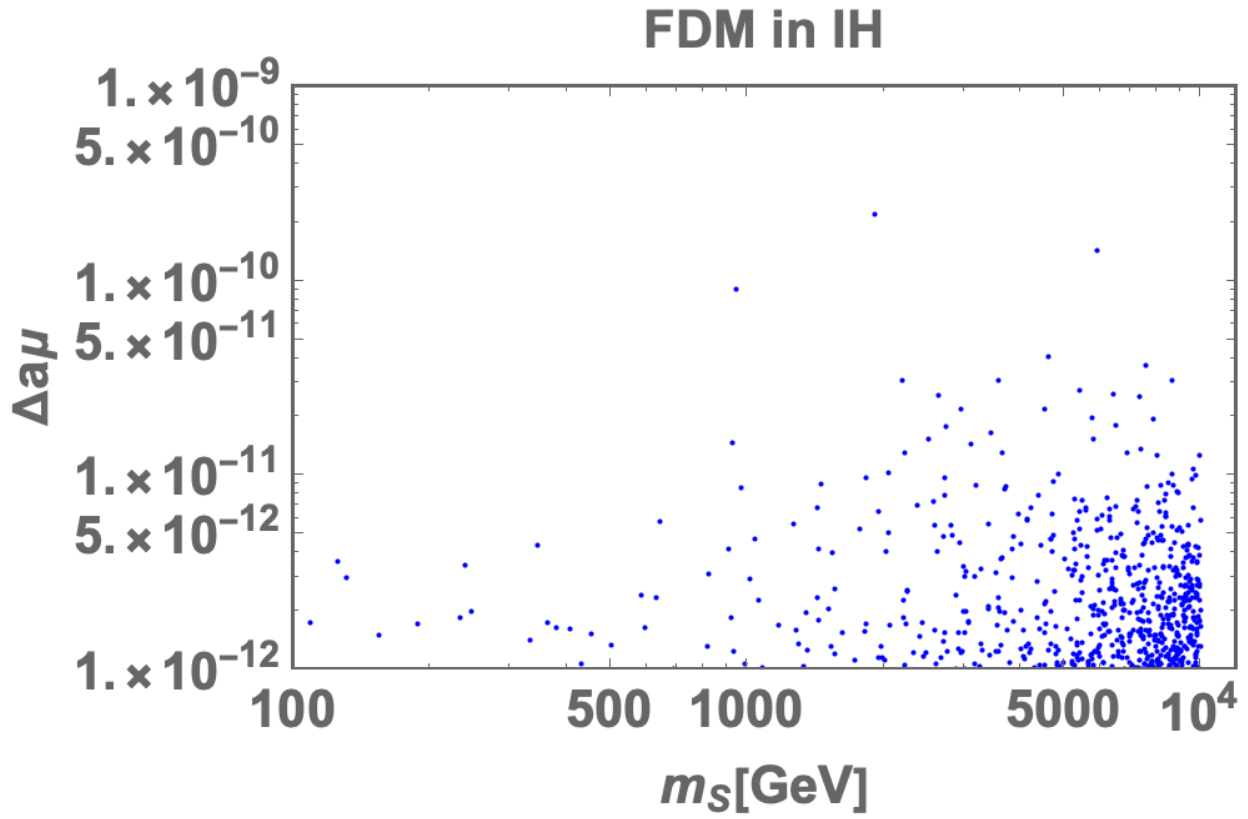}
\includegraphics[scale=0.22]{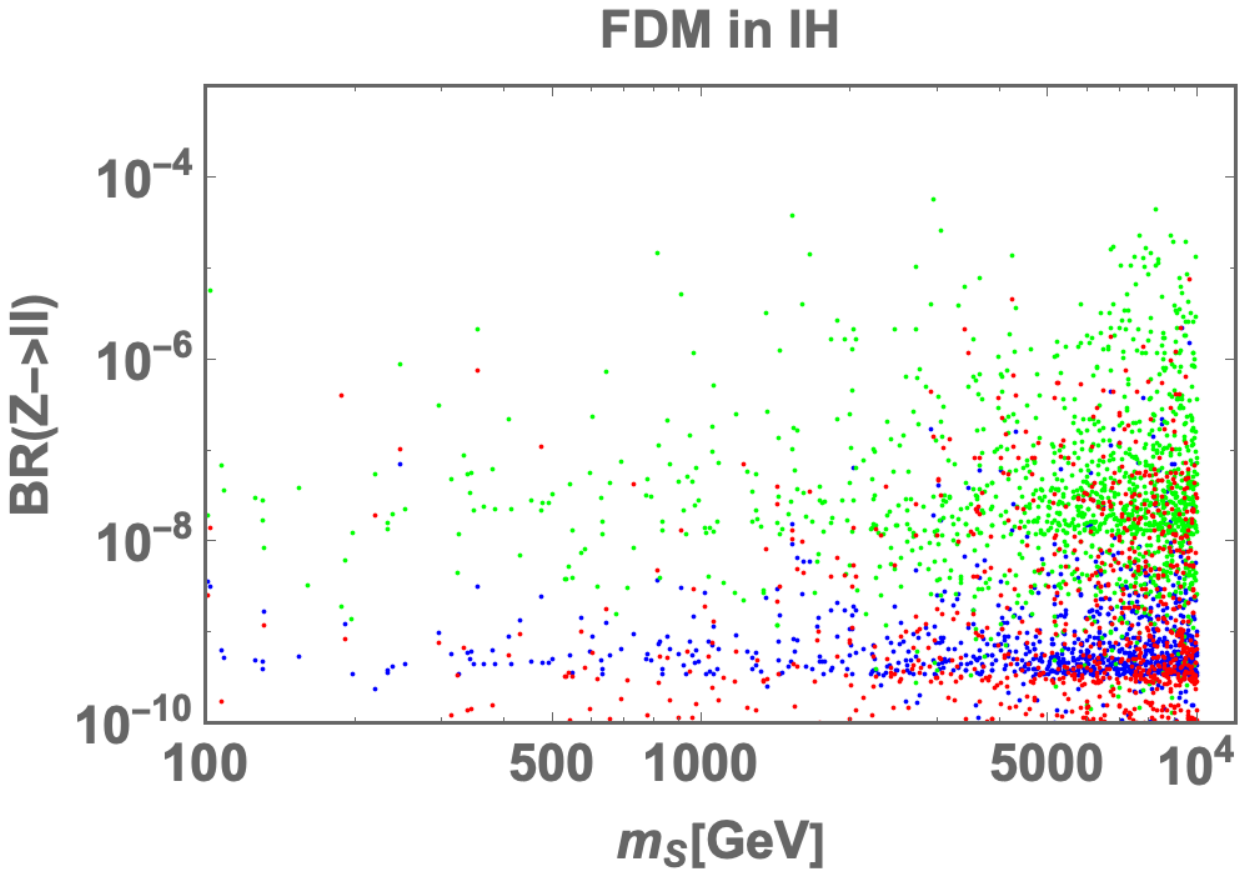}
\caption{Numerical results for the case of FDM candidate with IH where all the figure-legends are the same as the NH one. }
\label{fig:fdmih}
\end{center}
\end{figure}
%
In Figure~\ref{fig:fdmih}, we show numerical results for the case of FDM candidate with IH
where we have obtained 1,312 allowed plots to satisfy the LFVs and Z boson decays.
The left side figure represents the scattering plot of LFVs in terms of $m_S$ where the red points are for $\tau\to\mu\gamma$ and the blue ones $\mu\to e\gamma$.
The middle one represents the scattering plot of muon $g-2$ in terms of $m_S$. The order of muon $g-2$ reaches $10^{-10}$.
The right one shows the scattering plot of BR($Z\to\ell\bar\ell$) in terms of $m_S$ where ${\rm BR}(Z\to e\bar e)$ is blue, ${\rm BR}(Z\to \mu\bar\mu)$ green and ${\rm BR}((Z\to \tau\bar\tau))$ red.
The right one implies that many plots are located in places near each the experimental limit of this Z boson decays that would be well-tested by future experiments.
On the other hand, we obtained the maximum values of flavor non-conserving Z boson decays in our numerical analysis as follows:
\begin{align}
& {\rm BR}(Z \to e^\pm \mu^\mp)^{max} \approx 1.6 \times 10^{-10}, \quad {\rm BR}(Z \to e^\pm \tau^\mp)^{max} \approx 6.4 \times 10^{-7}, \nonumber \\
& {\rm BR}(Z \to \mu^\pm \tau^\mp)^{max} \approx 1.4 \times 10^{-7}.
\end{align}
Although ${\rm BR}(Z \to e^\pm \mu^\mp)$ and ${\rm BR}(Z \to \mu^\pm \tau^\mp)$ are much smaller than the experimental constraints, ${\rm BR}(Z \to e^\pm \tau^\mp)$ would be well-tested by future experiments.

\subsubsection{IH in case of BDM}
\begin{figure}[tb]
\begin{center}
\includegraphics[scale=0.22]{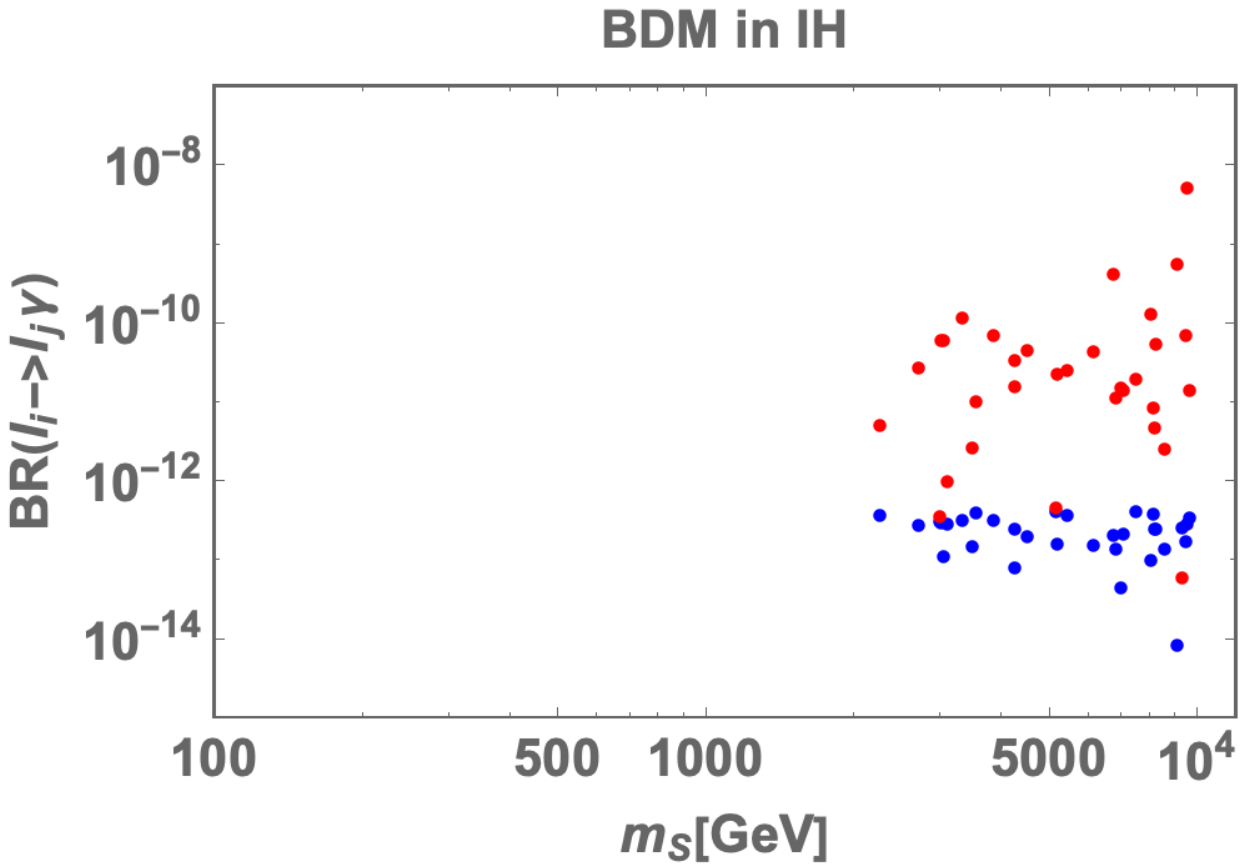}
\includegraphics[scale=0.22]{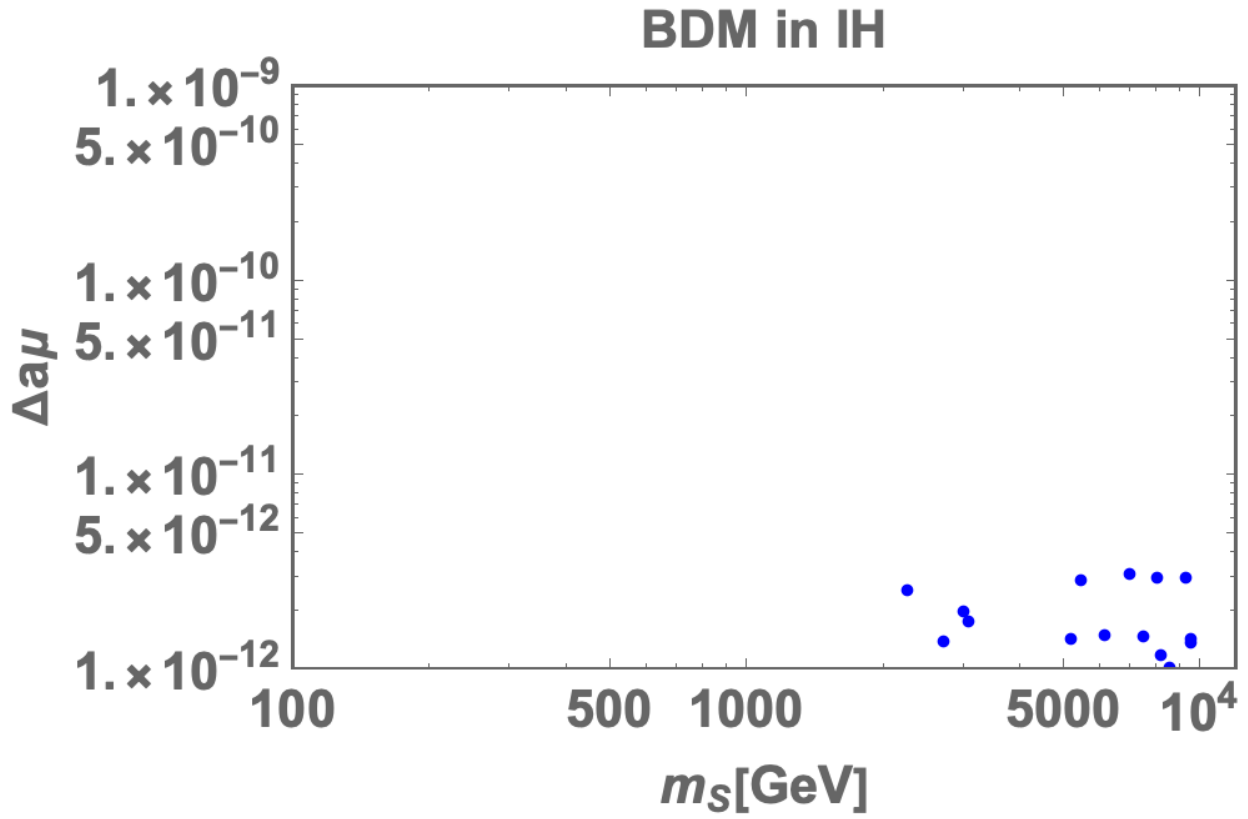}
\includegraphics[scale=0.22]{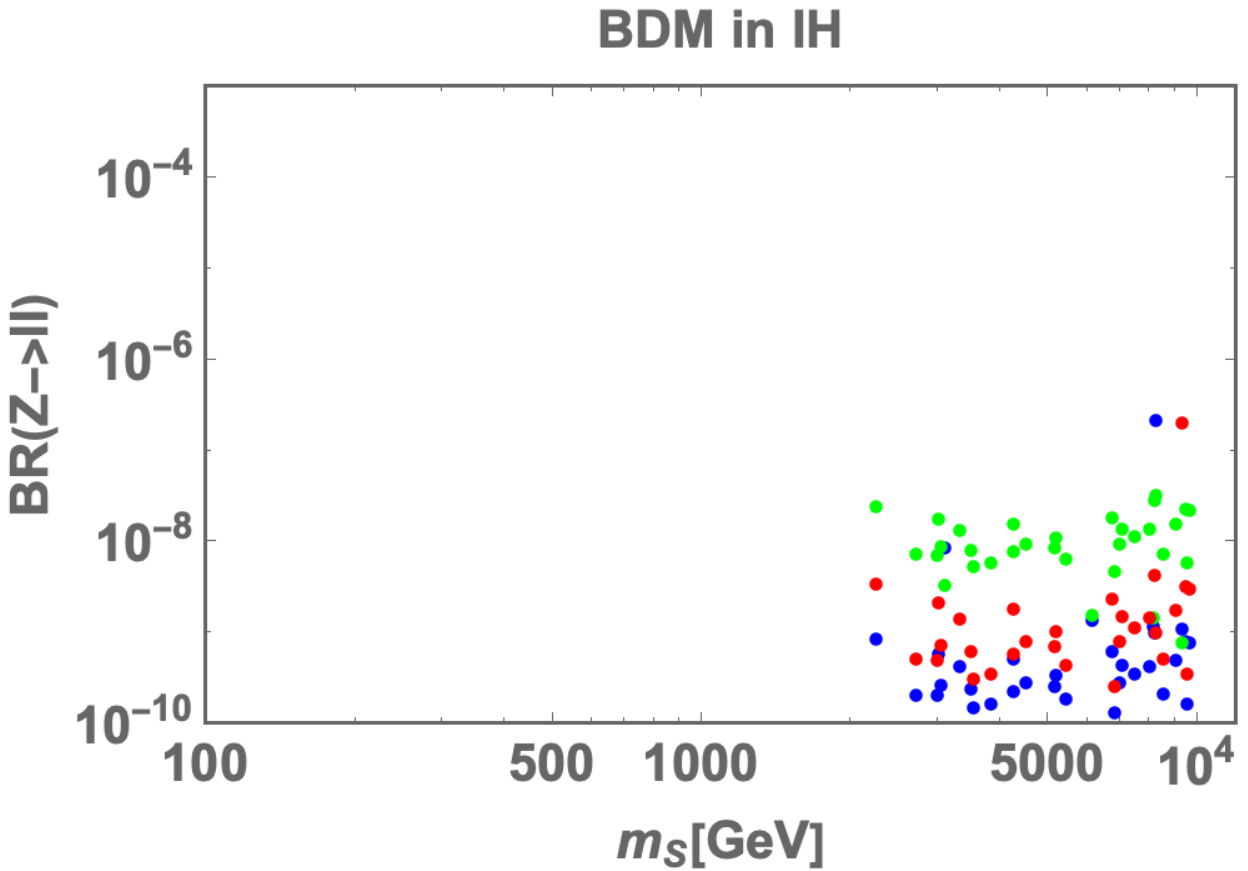}
\caption{Numerical results for the case of BDM candidate with IH where all the figure-legends are the same as the NH one.}
\label{fig:bdmih}
\end{center}
\end{figure}
%
In Figure~\ref{fig:bdmih}, we show numerical results for the case of BDM candidate with IH
where we have obtained 32 only allowed plots to satisfy the LFVs and Z boson decays.
The left side figure represents the scattering plot of LFVs in terms of $m_S$ where the red points are for $\tau\to\mu\gamma$ and the blue ones $\mu\to e\gamma$.
The middle one represents the scattering plot of muon $g-2$ in terms of $m_S$. The muon $g-2$ is less than $10^{-11}$ that is far from the expected value.
The right one shows the scattering plot of BR($Z\to\ell\bar\ell$) in terms of $m_S$ where ${\rm BR}(Z\to e\bar e)$ is blue, ${\rm BR}(Z\to \mu\bar\mu)$ green and ${\rm BR}((Z\to \tau\bar\tau))$ red.
All processes are less than these constraints by one order magnitude that would still be well-tested by future experiments.
On the other hand, we obtained the maximum values of flavor non-conserving Z boson decays in our numerical analysis as follows:
\begin{align}
& {\rm BR}(Z \to e^\pm \mu^\mp)^{max} \approx 1.0 \times 10^{-13}, \quad {\rm BR}(Z \to e^\pm \tau^\mp)^{max} \approx 3.6 \times 10^{-15}, \nonumber \\
& {\rm BR}(Z \to \mu^\pm \tau^\mp)^{max} \approx 1.9 \times 10^{-15},
\end{align}
which are much smaller than the experimental constraints.

\section{Dark Matter Particles}
In this section, we further propel our discussion on DM candidates FDM and its mass $D_{N_1}$ or BDM and its mass $m_{R}$, where we have just estimated their mass ordering among new particles in the neutrino and LFVs numerical analysis. 
{
The points used in the analysis of this section are those allowed by LFVs and derived in Subsection \ref{subsec:numericalanalysis}. 

The relic density of DM in the Universe is determined from observations to be 
\begin{align}
\Omega_{DM} h^2=0.120 \pm 0.001.
\label{eq:relicdensity}
\end{align}
In the thermal production scenario of DM, the relic abundance of DM is obtained by solving the Boltzmann equation \cite{Jungman:1995df}. 
It can be approximated by 
\begin{align}
\Omega_{DM} h^2 \simeq \frac{s_0}{\rho_c/h^2}\left(\frac{45}{\pi^2g_*^2} \right)^{1/2} \frac{m_{DM}}{T_f M_P \langle \sigma v\rangle},
\end{align}
where $s_0$ is the present entropy density of the Universe, $\rho_c$ is the critical density, $h$ is the scaled Hubble parameter, $g_*$ is the number of relativistic degrees of freedom when DM freezes out, $T_f$ is the freeze-out temperature, and $M_P$ is the Planck mass \cite{Baer:2014eja}. Thermally averaged DM annihilation cross section times DM relative velocity is denoted as $\langle \sigma v\rangle$.
The numerical analysis of the relic abundance is performed using \texttt{MicrOmegas 5.2.7} \cite{Giomataris:1995fq}.}

\subsection{ Fermion Dark Matter ($D_{N_1}$)}
We do not need to investigate constraints of direct detection bounds since FDM has no interactions with the quark sector at tree level.~\footnote{At one-loop level, there exist, but it would be easy to evade these bounds. See e.g. ~\cite{Abe:2018emu}.} 
Since the FDM interacts with SM particles only via Yukawa couplings given in Eq.(II.15), we might have a relatively clear result for analysis of the relic density of DM on whether we have solutions to satisfy the correct relic density.

{The thermally averaged cross section (to explain the relic density)
is given by
\begin{align}
&\langle\sigma v_{\rm rel}\rangle(2X\to\ell_i\bar\ell_j)
\approx  \frac1{16\pi s}
\int_0^\pi|\bar M|^2\sin\theta,\\
&|\bar M|^2=
\left((Y_S)_{1j}(Y_S^\dag)_{j1} + Y_{i1}Y^\dag_{1i}\right)
\left((Y_S)_{1i}(Y_S^\dag)_{i1}+Y_{j1}Y^\dag_{1j}\right)
\left[
\frac{(p_1\cdot k_1)(p_2\cdot k_2)}{(t-m_S^2)^2} 
+
\frac{(p_1\cdot k_2)(p_2\cdot k_1)}{(u-m_S^2)^2} 
\right]\nn\\
&-\frac{1}{t-m_S^2}\frac{1}{u-m_S^2}
\left[
\left(Y_{i1}Y^\dag_{1i}(Y_S)_{1j}(Y_S^\dag)_{j1}
+
Y_{j1}Y^\dag_{1j}(Y_S)_{1i}(Y_S^\dag)_{i1}
\right)(p_1\cdot k_1 p_2\cdot k_2+p_1\cdot k_2 p_2\cdot k_1
-p_1\cdot p_2 k_1\cdot k_2)\right.\nn\\
&\left.
+\left(Y_{i1}Y^\dag_{1i} Y_{1j} Y^\dag_{j1}
+
(Y_S)_{1i}(Y_S^\dag)_{i1}(Y_S)_{1j}(Y_S^\dag)_{j1}
\right) m_{DM}^2 (k_1\cdot k_2)
\right],
\end{align}
where we presume $m_{\ell_{i,j}}/m_S\approx0,\ m_{\ell_{i,j}}/m_{DM}\approx0$ in the above equations.
$p_1,\ p_2$ are four momenta of the DM initial states, 
$k_1$ and $k_2$ are respectively the ones of $\ell_i$ and $\bar\ell_j$ final states.

It can be approximately given by the s-wave contribution as a result in expansion of relative velocity and found as
\begin{align}
\langle\sigma v_{\rm rel}\rangle
\approx \frac{m_{DM}^2}{16\pi (m_{DM}^2+m_S^2)^2}
\left[(Y_S)_{1j}(Y_S^\dag)_{j1}Y_{i1}Y^\dag_{1i}+(Y_S)_{1i}(Y_S^\dag)_{i1}Y_{j1}Y^\dag_{1j}\right].
\end{align}
Roughly speaking, we need the cross section to be $10^{-9}$ GeV$^{-2}$ in order to satisfy the observed relic density.
However, our cross section is of the order $10^{-14}$ GeV$^{-2}$
 that is much smaller than the expected value.
 Thus, our FDM cannot be the main component of DM.
 }

\subsection{ Boson Dark Matter $m_R$}
\begin{figure}[tb]
\begin{center}
\includegraphics[scale=0.25]{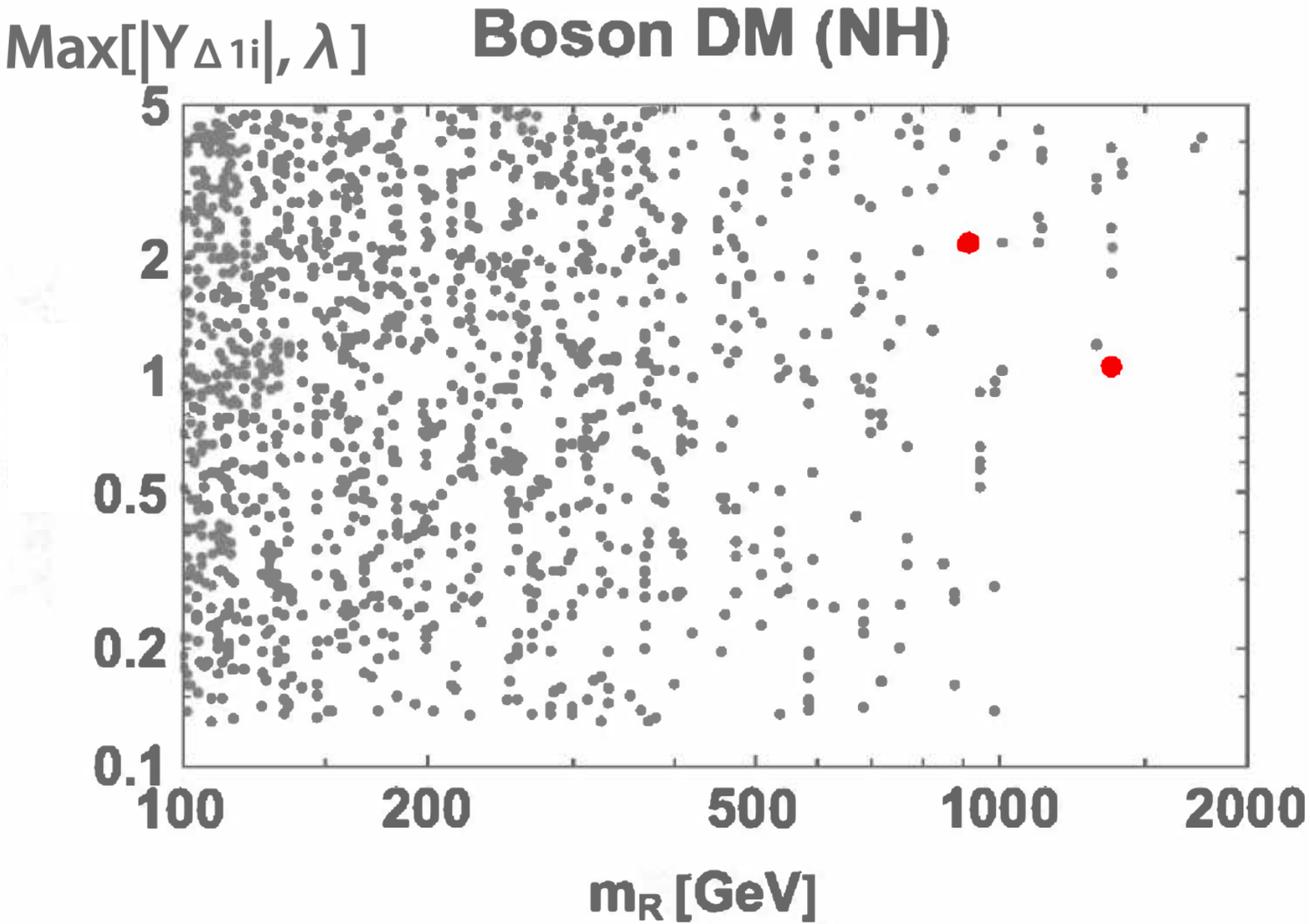}
\includegraphics[scale=0.25]{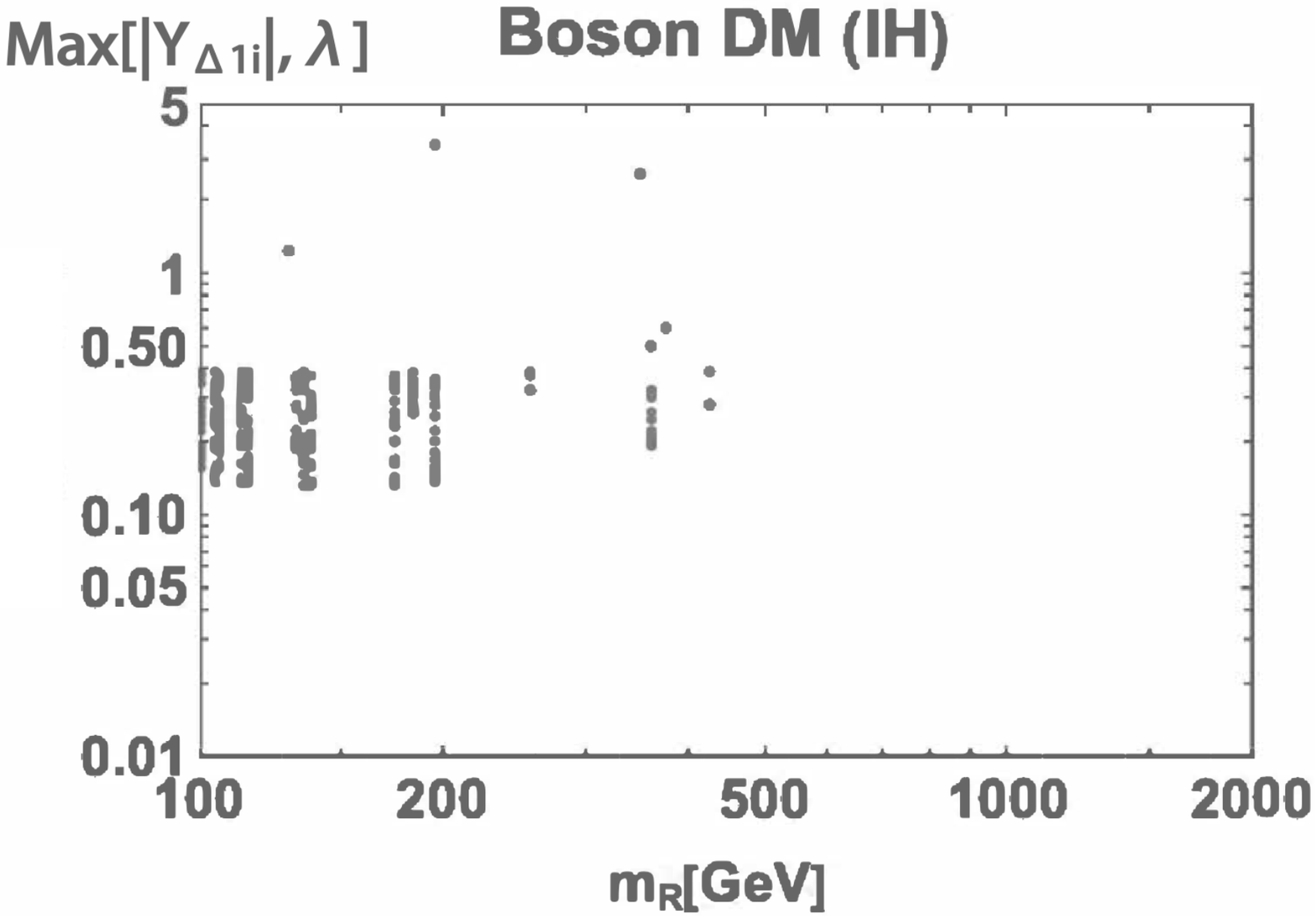}
\caption{The relic abundance of BDM for NH (left) and IH (right) cases. The vertical axis represents the largest absolute value of $y_{\Delta_{j1}}$, $\lambda_H$, $\lambda_S$, $\lambda_\Delta$, $\lambda_{HS}$, $\lambda_{H\Delta}$, and $\lambda'_{H\Delta}$. All points are allowed by LFVs.}
\label{fig:relic_BDM}
\end{center}
\end{figure}
In addition to the Yukawa interactions, BDM has potential interactions that always provide solutions via the resonant region of the SM Higgs or additional bosons. However, we need to consider the bounds on direct detection bounds since BDM always couples to the quark sector via Higgs potential~\cite{Kanemura:2010sh}.

{
Figure~\ref{fig:relic_BDM} shows the relic abundance for the case of BDM. 
The relic abundance tends to be less than observed one, and even though LFVs are satisfied. However, since $\Omega_{DM}h^2=10^{-8}-10^{-2}$ in many cases, few points agree with the observation. Especially in the IH case, the $m_R$ allowed by LFVs is small, roughly $m_R <$ 450 GeV. Since the relic abundance $\Omega_{DM} h^2\propto m_R$, it cannot be enhanced with a relatively small DM mass. In the parameter range, we scanned, no points consistent with the observation are found. 
Here, the main annihilation processes for the calculation of the relic abundance are gauge interactions, e.g., ${\delta^+, \delta^-} \to W^+, W^-$
and $\delta_I, \delta_I \to W^+, W^-$. Note that since $m_R$ is degenerate with the masses of the same scalar sector {$m_{\delta^+}, m_{\delta^-}$}
and $m_I$, their coannihilation processes contribute to the calculation. The scalar potential interactions like ${\delta^+, \delta^-} \to h h$
only work secondarily, so the effects by $\lambda$ couplings are insignificant. 
Annihilation from $\delta_{R,I}$ to SM leptons can also occur through $y_{\Delta_{ij}}$, but this contribution is even smaller since $|\lambda| \gg |y_{\Delta_{ij}}|$ in many cases.
}

\section{ Conclusions and discussions}
We have studied a one-loop induced radiative neutrino model with an inert isospin triplet boson of zero hypercharge, in which we have discussed
neutrino oscillations, lepton flavor violations, Z boson decays, muon anomalous magnetic dipole moment, and shown the allowed region of our parameter space in both the cases for normal hierarchy and inverted one.
When we do not consider the dark candidate, We have obtained sizable scale of 
muon anomalous magnetic dipole moment
that is  $10^{-9}$ for both cases.
Then, we have discussed fermionic dark matter 
and bosonic dark matter 
candidates to explain the relic density of dark matter in light of our numerical analysis of the neutrino oscillation and lepton flavor violations. 

We have found muon $g-2$ reaches 
the order $10^{-9}$ in cases of normal hierarchy and inverted one when we do not consider the dark matter candidate. 
When we consider the dark matter candidate, we have found it to be the order $10^{-10}$ for the case of fermionic dark matter candidate even though the dark matter cannot be the main component of dark matter in analysis of relic density. 
Although the flavor non-conserving processes of Z boson decays tend to be much smaller than the current bounds, the conserving ones reach the current bounds which would be well-tested in future experiments.

In our dark matter analysis, we have found that our fermionic dark matter candidate would not be the main component of the dark matter since the corresponding cross section is too small. But, we have found bosonic dark matter can satisfy the observed relic density.

\section*{Acknowledgments}
 This research was supported by an appointment to the YST
Program at the APCTP through the Science and Technology Promotion Fund and Lottery Fund of the Korean Government.
This was also supported by the Korean Local Governments - Gyeongsangbuk-do Province and Pohang city (S.J.). 
The work was also supported by JSPS Grant-in-Aid for Scientific Research (C) 21K03562 and (C) 21K03583 (K. I. N). 
\vspace{0.5cm}


\section*{ Appendix}
Here we explicitly show four momentum inner products for the process $p_1(m_1),p_2(m_2)\to k_1(n_1),k_2(n_2)$ as functions of their masses and integrated parameters $\theta$ and one of the Mandelstam valuables $s$
\begin{align}
 p_1\cdot p_2 &=\frac{s-m_1^2 - m_2^2}{2}\;,\quad k_1\cdot k_2=\frac{s-n_1^2 - n_2^2}{2}
\;,
\end{align}
\begin{align}
p_1\cdot k_1 &=\frac{|(s+m_1^2-m_2^2)(s+n_1^2-n_2^2)|-
\cos\theta\sqrt{[(s-m_1^2-m_2^2)^2-4m_1^2m_2^2][(s-n_1^2-n_2^2)^2-n_1^2n_2^2]}}
{4 s}
\;,
\end{align}
\begin{align}
p_1\cdot k_2 &=\frac{|(s+m_1^2-m_2^2)(s-n_1^2+n_2^2)|+\cos\theta\sqrt{[(s-m_1^2-m_2^2)^2-4m_1^2m_2^2][(s-n_1^2-n_2^2)^2-n_1^2n_2^2]}} {4 s} 
\;,
\end{align}
\begin{align}
p_2\cdot k_1 &= \frac{|(s-m_1^2+m_2^2)(s+n_1^2-n_2^2)|+\cos\theta\sqrt{[(s-m_1^2-m_2^2)^2-4m_1^2m_2^2][(s-n_1^2-n_2^2)^2-n_1^2n_2^2]}} {4 s} 
\;,
\end{align}
\begin{align}
p_2\cdot k_2 &=\frac{|(s-m_1^2+m_2^2)(s-n_1^2+n_2^2)|-\cos\theta\sqrt{[(s-m_1^2-m_2^2)^2-4m_1^2m_2^2][(s-n_1^2-n_2^2)^2-n_1^2n_2^2]}} {4 s} ,
\end{align}
$s$ can be expanded them in term of $v_{\rm rel}$;
$s=(m_1+m_2)^2 +m_1m_2 v^2_{\rm rel}$, that is a good approximation if there exist any poles in the cross section.

\bibliography{sno1_revise.bib}

\end{document}